\documentstyle[12pt,epsf]{article}

\catcode `\@=11 \@addtoreset{equation}{section}

\newcommand{\plot}[1]
{\begin{center}
\epsfysize=6.5cm 
\vspace{-5mm}
\parbox{\epsfxsize}{\epsffile{#1}}
\vspace{5mm}
\end{center}}

\def\be{\begin{equation}}
\def\ee{\end{equation}}
\def\l{\label}
\def\dx{\frac{d}{dx}}
\def\dy{\frac{d}{dy}}
\def\ER{\frac{MER^2}{2}}
\def\eV{\mbox{ eV}}
\def\au{\mbox{ a.u.}}
\def\bohr{\mbox{ bohr}}
\def\hartree{\mbox{ hartree}}

\def\lm{\lambda }
\begin{document}

\begin{titlepage}
\title{\Large\bf EXACT SOLUTION OF THE RESTRICTED THREE-BODY 
SANTILLI-SHILLADY MODEL OF $H_2$ MOLECULE}
\author{{\normalsize\bf A.K.Aringazin$^1$ and M.G.Kucherenko$^2$}\\[0.3cm]
{\normalsize $^1$Karaganda State University, Karaganda 470074 Kazakstan}\\
{\normalsize ascar@ibr.kargu.krg.kz}\\
{\normalsize $^2$Orenburg State University, 13 Pobedy Ave., Orenburg
460352 Russia}\\{\normalsize rphys@osu.ac.ru}\\}
\date{December 1999}
\maketitle
\abstract{
In this paper, we study the exact solution of the restricted isochemical 
model of $H_2$ molecule with fixed nuclei recently proposed by 
Santilli and Shillady in which the two electrons are assumed to be 
bonded/correlated into a quasiparticle called the {\it isoelectronium}. 
Under the conditions that: 1) the isoelectronium is stable; 2) the effective 
size of the isoelectronium is ignorable, in comparison to internuclear 
distance; and 3) the two nuclei are at rest, the Santilli-Shillady model 
of the $H_2$ molecule is reduced to a restricted {\it three-body} system 
essentially similar to a neutral version of the $H_2^+$ ion, which, as such, 
admits exact solution.
Our main result is that the restricted three-body Santilli-Shillady approach 
to $H_2$ is capable to fit the experimental binding energy, 
at the isoelectronium mass parameter $M=0.308381m_e$, although under 
optimal internuclear distance about 19.6\% bigger than the conventional 
experimental value, indicating an approximate character of the three-body 
model.}
\end{titlepage}
\newpage

\section{Introduction}

In this paper, we study isochemical model of the $H_2$ molecule recently 
introduced by R. M. Santilli and D. D. Shillady \cite{SS}, which is 
characterized by the conventional $H_2$ model set up plus a short-range 
attractive Hulten potential interaction between the two electrons originating 
from the deep overlapping of their wave functions at mutual distances of the 
order of 1 fm;  see also \cite{SS2}. If one assumes that this attractive potential
is strong enough to overcome Coloumb repulsion between the two electrons, 
they can form electron-electron system called {\it isoelectronium}. 
The isoelectronium is characterized by "bare" mass $M=2m_e$, 
as a sum of masses of two constituent electrons, charge $-2e$, radius about 
$10^{-11}$ cm, and null magnetic moment. The used Hulten potential contains 
two real parameters, one of which is the isoelectronium correlation length 
parameter $r_c$, which can be treated as an effective radius of isoelectronium.

The main structural difference between the Santilli-Shillady isochemical model 
and the conventional quantum chemical model of the $H_2$ molecule, is that the 
former admits additional nonlinear, nonlocal, and nonpotential, thus nonunitary 
effects due to the deep overlapping of the wavepackets of valence electrons at 
short distances, which are responsible for the strong molecular bond. 
In a first nonrelativistic approximation, Santilli and Shillady \cite{SS} 
derived the following characteristics of the isoelectronium:
total rest mass $M = 2m_e$, charge $-2e$, magnetic moment zero,
and radius $6.84323\times 10^{-11} cm$. The value $M=2m_e$ of
the rest mass was derived via the assumption of a contact, {\it nonpotential} 
interactions due to the mutual wave-overlapping sufficiently strong to 
overcome the repulsive Coulomb force. The nonpotential character of the bond
was then responsible for the essential lack of binding energy in
the isoelectronium, and the resulting value $M=2m_e$.
However, the authors stressed in \cite{SS} that the isoelectronium
is expected to have a non-null binding energy, and, therefore, a rest mass 
smaller than $2m_e$. One argument presented in \cite{SS} is that, when
coupled in singlet at very short distances, the two electrons eventually experience
very strong {\it attractive} forces of magnetic type, due to
the two pairs of opposing magnetic polarities, resulting in a bond. 
The potential origin of the bond then implies the existence of a binding energy, 
resulting in a rest mass of the isoelectronium smaller than $2m_e$.
Also, in the subsequent paper \cite{Santilli}, Santilli pointed out that
the isoelectronium can at most admit a small instability. 

As a result of a correlation/bonding between the two electrons, 
Santilli and Shillady were able to reach, for the first time, 
representations of the binding energy and other characteristics 
of $H_2$ molecule which are accurate to the {\it seventh digit}, 
within the framework of numerical Hartree-Fock approach to $H_2$ 
molecule viewed as a four-body system with fixed nuclei, and with 
the use of Gaussian screened Coloumb potential taken as an approximation 
to the Hulten potential \cite{SS}.

On the other hand, the above mentioned strong short-range character 
of the electron-electron interaction suggests the use of approximation 
of {\it stable} isoelectronium of {\it ignorably small} size, in comparison 
to the internuclear distance \cite{SS}.
Indeed, under these two assumptions one can reduce the conventional 
{\it four-body} structure of the $H_2$ molecule to a {\it three-body} 
system (the two electrons are viewed as a single point-like particle).
Furthermore, in the Born-Oppenheimer approximation, i.e. at fixed 
nuclei, we have a {\it restricted} three-body system, the Shr\"odinger 
equation for which admits {\it exact} analytic solution. 

So, we have the original four-body Santilli-Shillady model of $H_2$
molecule, and the three-body Santilli-Shillady model of $H_2$,
which is an approximation to it. The former is characterized by, 
in general, unstable isoelectronium and, thus, sensitivity to details 
of the electron-electron interaction, while the latter deals with 
a single point-like particle (stable isoelectronium of ignorable size) 
moving around two fixed nuclei.

Clearly, the three-body Santilli-Shillady model of $H_2$ molecule
can be viewed as {$H_2^+$ ion like system}.
For the sake of brevity and to avoid confusion with the $H_2^+$ ion itself, 
we denote $H_2$ molecule, viewed as the restricted three-body system, 
as $\hat H_2$. Note that $\hat H_2$ is a neutral $H_2^+$ ion like system.

The quantum mechanical problem of the restricted $H_2^+$ ion like systems, 
associated differential equation, and its exact analytic solution have been 
studied in the literature by various authors \cite{Landau}-\cite{Wind}.

In this paper we present the exact analytic solution of the above 
indicated restricted three-body Santilli-Shillady isochemical
model of the hydrogen molecule, study its asymptotic behavior, and analyze 
the ground state energy, presenting numerical results in the form of tables
and plots. Our analysis is based on the analytical results obtained for 
thoroughly studied $H_2^+$ ion.

In Sec.~\ref{SecSS}, we review some features of the four-body 
Santilli-Shillady model of $H_2$ necessary for our study, 
and introduce our separation of variables in the Schr\"odinger 
equation under the assumption that the isoelectronium is a stable 
quasiparticle of ignorable size.

In Sec.~\ref{H2+}, we review the exact analytic solution of the 
$H_2^+$ ion like systems (which includes the $\hat H_2$ system), and
study their asymptotic behavior at large and small distances between 
the two nuclei.

In Sec.~\ref{Scaling}, we use the preceding solution to find the binding 
energy of $\hat H_2$ system. We then develop a scaling method and use 
Ritz's variational approach to check the results. Both the cases of the 
isoelectronium "bare" mass $M=2m_e$ and of variable mass parameter, 
$M=\eta m_e$ have been studied. All the data and basic results of this 
Section have been collected in Table~\ref{Table1}.

In Sec.~\ref{Concluding}, we introduce a preliminary study on
the application of Ritz's variational approach to the general four-body 
Santilli-Shillady model of $H_2$, where the isoelectronium is an 
{\it unstable} composite particle, in which case the model re-acquires 
its {\it four-body} structure, yet preserves a strong bonding/correlation 
between the electrons.

In the Appendix, we present the results of our numerical calculations
of the ground state energy of $H_2^+$ ion and of $\hat H_2$ system, 
for different values of the isoelectronium mass parameter $M$, 
based on their respective exact solutions, in the form of tables and plots.

Our main result is that {\it the restricted three-body Santilli-Shillady 
isochemical model of the hydrogen molecule does admit exact analytic 
solution capable of an essentially exact representation of the binding 
energy, although under internuclear distance about 19.6\% bigger than 
the conventional experimental value.}
The mass parameter $M$ of isoelectronium has been used here to fit the 
experimental value of the binding energy, with the result 
$M= 0.308381m_e$ (i.e. about 7 times less than the "bare" mass $M=2m_e$).
In this paper, we assume that some defect of mass effect may have 
place leading to decrease of the "bare" mass $M=2m_e$. 

We also note that the value $M = 0.308381 m_e$ implies
a binding energy of about $1.7$ MeV, which is admittedly rather large.
Recent studies by Y. Rui \cite{Rui} on the correct force law among 
spinning charges have indicated the existence of a critical 
distance below which particles with the same  charge attract each others. 
If confirmed, these studies imply that the repulsive Coulomb force itself 
between two electrons in singlet coupling  can be attractive
at a sufficiently small distance, thus eliminating the need to postulate 
an attractive force sufficiently strong to overcome the repulsive 
Coulomb force. As a result, a binding energy in the isoelectronium structure
of the order of 1.7 MeV cannot be excluded on grounds of our knowledge 
at this time.

Clearly, however, that due to the current lack of dynamical description
of the above mentioned defect of mass, and the obtained result that the 
predicted internuclear distance is about 19.6\% bigger than the experimental 
value, our study is insufficient to conclude that the isoelectronium is 
permanently stable, and one needs for additional study on the four-body 
Santilli-Shillady isochemical model of $H_2$, which is conducted in 
a subsequent paper by one of the authors \cite{Aringazin}.

\section{Santilli-Shillady model of $H_2$ molecule}
\l{SecSS}
\subsection{General equation}

The Santilli-Shillady iso-Shr\"odinger's equation for $H_2$ molecule with
short-range attractive Hulten potential between the two electrons can be 
reduced to the following form \cite{SS}:
\begin{eqnarray}\l{general}
\left( -\frac{\hbar^2}{2m_1}\nabla^2_1
-\frac{\hbar^2}{2m_2}\nabla^2_2
-V_0\frac{e^{-r_{12}/r_c}}{1-e^{-r_{12}/r_c}} +\frac{e^2}{r_{12}}
\right.
\\ \nonumber
\left.
-\frac{e^2}{r_{1a}}  -\frac{e^2}{r_{2a}}
-\frac{e^2}{r_{1b}}  -\frac{e^2}{r_{2b}}
+\frac{e^2}{R}
\right)|\phi\rangle = E|\phi\rangle,
\end{eqnarray}
where $V_0$ and $r_c$ are positive constants, and $R$ is
distance between nuclei $a$ and $b$.
By using vectors of center-of-mass system of electrons 1 and 2,
$\vec{r}_a$ and $\vec{r}_b$, originated at nuclei $a$  and $b$,
respectively, we have
\be\l{|A|}
r_{1a}=\left|\vec{r}_a - {m_2\over m_1+m_2}\vec{r}_{12}\right|,
\quad
r_{2a}=\left|\vec{r}_a + {m_1\over m_1+m_2}\vec{r}_{12}\right|.
\ee
\be\l{|B|}
r_{1b}=\left|\vec{r}_b - {m_2\over m_1+m_2}\vec{r}_{12}\right|,
\quad
r_{2b}=\left|\vec{r}_b + {m_1\over m_1+m_2}\vec{r}_{12}\right|,
\ee
(for electrons we have $m_1=m_2=m_e$). The Lagrangian of the system
can be written
\be
{\cal L}={m_1\dot{r}_1^2\over 2}+ {m_2\dot{r}_2^2\over 2} -
V(r_{12}) - W(r_{1a},r_{1b},r_{2a},r_{2b},R), \label{L}
\ee
Here, $V$ is the potential energy of interaction between the electrons 
1 and 2,
\be\l{U}
V(r_{12})= \frac{e^2}{r_{12}}-V_0\frac{e^{-r_{12}/r_c}}{1-e^{-r_{12}/r_c}},
\ee
and $W$ is the potential energy of interaction between electrons and nuclei,
and between two nuclei,
\be\label{W}
W(r_{1a},r_{1b},r_{2a},r_{2b},R)=
-\frac{e^2}{r_{1a}} -\frac{e^2}{r_{2a}}
-\frac{e^2}{r_{1b}} -\frac{e^2}{r_{2b}}+\frac{e^2}{R}.
\ee
Notice that $\dot{r}_1 = \dot{r}_{1a}=\dot{r}_{1b}$,
and $\dot{r}_2 = \dot{r}_{2a}=\dot{r}_{2b}$,
because $\vec{r}_{1a}=\vec{r}_{1b}+\vec{R}$
and $\vec{r}_{2a}=\vec{r}_{2b}+\vec{R}$,
where $\vec R$ is constant vector. Similarly,
\be
\vec{r}_a=\vec{r}_b+\vec{R}, \quad
\vec{r}_a={m_1\vec{r}_{1a}+m_2\vec{r}_{2a}\over m_1+m_2},\quad
\vec{r}_b={m_1\vec{r}_{1b}+m_2\vec{r}_{2b}\over m_1+m_2}.
\ee
Then, Lagrangian (\ref{L}) can be rewritten as
${\cal L}={\cal L} (r_a, r_b, r_{12})$,
\be\l{L'}
{\cal L}={M\dot{r}_a^2\over 2}+{m\dot{r}_{12}^2\over 2} - V(r_{12})
- W(r_{a},r_{b},r_{12},R).
\ee
Here, $M=m_1+m_2$  is the total mass of the electrons, and
$m = m_1m_2/(m_1+m_2)$ is the reduced mass.
Corresponding generalized momenta take the form
\be
\vec{P}_M={\partial{\cal L}\over \partial\dot{\vec{r}_A}}=M\dot{\vec{r}_A}.
\quad \vec{p}_{m}={\partial{\cal L}\over \partial\dot{\vec{r}_{12}}}
=m\dot{\vec{r}_{12}}.
\ee
The system reveals axial symmetry, with the axis
connecting two nuclei. Also, for identical nuclei we have
reflection symmetry in respect to the plane perpendicular to
the above axis and lying on equal distances from the two nuclei.

\subsection{Separation of variables}

Santilli and Shillady \cite{SS} then assume that,
as a particular case under study in this paper 
(not to be confused with the general four-body case), 
the two valence electrons of the $H_2$ molecule can form a 
stable quasi-particle of small size due to short-range attractive 
Hulten potential, such that
\be\l{approx}
r_{12}\ll r_a, \quad
r_{12}\ll r_b.
\ee
Therefore, we can ignore $r_{12}$ in Eqs.(\ref{|A|}) and (\ref{|B|}),
\be
\vec{r}_{1a}\approx\vec{r}_{2a}\approx\vec{r}_{a}, \quad
\vec{r}_{1b}\approx\vec{r}_{2b}\approx\vec{r}_{b}.
\ee
The Hamiltonian of the system then becomes
\be\l{H}
\hat{H}={\hat{P}_M^2\over 2M} + {\hat{p}_{m}^2\over 2m } +
V(r_{12}) + W(r_a, r_b, R),
\ee
where
\be
W(r_a, r_b, R)= - \frac{2e^2}{r_{a}}
-  \frac{2e^2}{r_{b}} + \frac{2e^2}{R}.
\ee
In this approximation, it is possible to {\it separate} the
variables $r_{a,b}$ and $r_{12}$. Namely,
inserting $|\phi\rangle =\psi (r_a, r_b, R)\chi(r_{12})$
into the equation \cite{SS}
\be\l{EqS}
\left( -\frac{\hbar^2}{2M}\nabla^2_{ab}
-\frac{\hbar^2}{2m}\nabla^2_{12}
-V_0\frac{e^{-r_{12}/r_c}}{1-e^{-r_{12}/r_c}}
+\frac{e^2}{r_{12}}
-\frac{2e^2}{r_{a}}  -\frac{2e^2}{r_{b}}
+\frac{e^2}{R}
\right)|\phi\rangle = E|\phi\rangle
\ee
we obtain
\be\l{Sep}
-\frac{\hbar^2}{2M}{\nabla^2_{ab}\psi\over\psi}
-\frac{\hbar^2}{2m}{\nabla^2_{12}\chi\over\chi} +V(r_{12})
+ W(r_a, r_b, R)-E=0.
\ee
By separating the variables, we have the following two
equations:
\be\l{r12}
-\frac{\hbar^2}{2m}{\nabla^2_{12}\chi}+ V(r_{12})\chi
= \varepsilon\chi.
\ee
and
\be\l{C}
-\frac{\hbar^2}{2M}{\nabla^2_{ab}\psi}+W(r_a, r_b, R)\psi
=(E-\varepsilon )\psi.
\ee
In this way, under approximation (\ref{approx}), the original 
four-body problem is reduced to a three-body problem
characterized by two differential equations:

1) Equation (\ref{r12}), which describes the electron-electron system 
forming the bound quasi-particle state called isoelectronium,
with "bare" total mass $M=2m_e$ and charge $-2e$.
This equation will not be studied in this paper, since we assume that the
isoelectronium is permanently stable.

2) Equation (\ref{C}), which is the structural equation of the
restricted three-body Santilli-Shillady isochemical model $\hat H_2$, 
in which the stable isoelectronium with "bare" mass $M=2m_e$, charge $-2e$, 
null magnetic moment and ignorable size orbits
around the two nuclei, hereon assumed to have infinite mass
(the Born-Oppenheimer approximation).

This paper is devoted to the study of the exact analytic solution 
of the latter equation, and its capability to represent the experimental 
data on the binding energy, bond length, and other characteristics of the
hydrogen molecule.

\section{Exact solution for $H_2^+$ ion like system}\l{H2+}

In this Section, we present analytical solution of the
Schr\"odinger equation for $H_2^+$ ion-like systems
in Born-Oppenheimer approximation, we
analyze the associated recurrence relations, and asymptotic behavior
of the solutions at large and small distances between the two nuclei.
As it was indicated \cite{SS},
this problem arises when Santilli-Shillady model
of $H_2$ is reduced to the restricted {\it three-body} problem
characterized by Eq. (\ref{C}), which possesses exact solution 
under appropriate separation of variables.

\subsection{Differential equations}
In Born-Oppenheimer approximation, i.e., at fixed nuclei,
the equation for $H_2^+$ ion-like system for a particle
of mass $M$ and charge $q$ is
\be\l{eqH2plus}
\nabla^2\psi + 2M(E+\frac{q}{r_a}+\frac{q}{r_b})\psi=0.
\ee
In spheroidal coordinates,
\be
x=\frac{r_a+r_b}{R}, \quad 1<x<\infty,
\ee
\be
y=\frac{r_a-r_b}{R}, \quad -1<y<1,
\ee
\be
\varphi, \quad  0<\varphi<2\pi,
\ee
where $R$ is a fixed separation distance between the nuclei $a$ and $b$,
and
\begin{eqnarray}
\nabla^2
=\frac{4}{R^2(x^2-y^2)}
\left(
\frac{\partial}{\partial x}(x^2-1)\frac{\partial}{\partial x}
+\frac{\partial}{\partial y}(1-y^2)\frac{\partial}{\partial y}
\right)
\\ \nonumber
+\frac{1}{R^2(x^2-1)(1-y^2)}\frac{\partial^2}{\partial\varphi^2}.
\end{eqnarray}
We then have from Eq.(\ref{eqH2plus})
\begin{eqnarray}\l{eqH2plus2}
\left[
\frac{\partial}{\partial x}(x^2-1)\frac{\partial}{\partial x}
+\frac{\partial}{\partial y}(1-y^2)\frac{\partial}{\partial y}
+\frac{x^2-y^2}{4(x^2-1)(1-y^2)}\frac{\partial^2}{\partial\varphi^2}
\right.
\\ \nonumber
\left.
+\frac{MER^2}{2}(x^2-y^2) + 2MqRx \right]\psi =0.
\end{eqnarray}
Here, we have used
\be
\frac{1}{r_a}+\frac{1}{r_b} = \frac{4}{R}\frac{x}{x^2-y^2}.
\ee
Obviously, Equation (\ref{eqH2plus2}) can be {\it separated} 
by the use of the representation
\be
\psi=f(x)g(y)e^{im\varphi},
\ee
under which we have two second-order {\it ordinary} differential equations,
\be\l{f}
\dx\left((x^2-1)\dx f\right)
-\left(\lm - 2MqRx - \ER x^2 + \frac{m^2}{x^2-1}\right)f=0,
\ee
\be\l{g}
\dy\left((1-y^2)\dy g\right)
+\left(\lm - \ER y^2 - \frac{m^2}{1-y^2}\right)g=0,
\ee
where $\lm$ is a separation constant (cf. \cite{Landau}).
So, the problem is to identify solutions for $f$ and $g$.

\subsection{Recurrence relations}

By introducing the re-formulations
\be f \to (x^2-1)^{m/2}f, \ee
\be g \to (1-y^2)^{m/2}g, \ee
to handle singularities at $x=\pm1$ and $y=\pm1$
in Eqs.(\ref{f}) and (\ref{g}), respectively,
we reach the following final form of the equations
to be solved:
\be\l{f''}
(x^2-1)f'' + 2(m+1)f' -(\lm + m(m+1) - \tilde a x - c^2x^2)f=0
\ee
and
\be\l{g''}
(1-y^2)g'' - 2(m+1)g' +(\lm - m(m+1) - c^2y^2)g=0,
\ee
where we have denoted
\be\l{ca}
c^2=\ER, \quad \tilde a= 2MqR.
\ee
We shall look for solutions in the form of power series.
Substituting the power series
\be f = \sum f_kx^k, \ee
\be g = \sum g_ky^k, \ee
into Eqs. (\ref{f''}) and (\ref{g''}), we obtain the recurrence relations,
\be\l{fn}
c^2f_{n-2} + \tilde a f_{n-1} - (\lm-(m+n)(m+n+1))f_n
\ee
$$
- (n+1)(n+2)f_{n+2}=0
$$
and
\be\l{gn}
c^2g_{n-2} - (\lm-(m+n)(m+n+1))g_n - (n+1)(n+2)g_{n+2}=0,
\ee
from which coefficients $f_k$ and $g_k$ must be found.
Here, $f_0$ and $g_0$ are fixed by normalization of the general
solution. Note that the recurrence relation (\ref{fn}) contains term
$2MqRf_{n-1}$ raised from the linear term $2MqRx$ in Eq.(\ref{f}).

In the next two Sections we consider some particular cases of interest
prior to going into details of the general solution. These particular
solutions are important for the study of the general case.

\subsection{The particular case $R=0$}\l{R0}

In the particular case $R=0$, the two nuclei are superimposed,
so that the system is reduced to a helium-like system,
\be\l{f2}
\dx\left((x^2-1)\dx f\right)
-\left(\lm+\frac{m^2}{x^2-1}\right)f=0,
\ee
\be\l{g2}
\dy\left((1-y^2)\dy g\right)
+\left(\lm-\frac{m^2}{1-y^2}\right)g=0.
\ee
From recurrence relations (\ref{fn}) and (\ref{gn})
we obtain the following particular recurrence sequences,
\be\l{fn2}
(\lm-(m+n)(m+n+1))f_n - (n+1)(n+2)f_{n+2}=0
\ee
and
\be\l{gn2}
(\lm-(m+n)(m+n+1))g_n - (n+1)(n+2)g_{n+2}=0,
\ee
which are equivalent to each other,
and can be {\it stopped} by putting the separation constant
\be \lm = (m+n)(m+n+1) = l(l+1), \ee
with $m= -l, \dots, l.$
This gives us well known solution for $g$ in terms of
Legendre polynomials,
\be g = (1-y^2)^{m/2}\frac{d}{dy^m}P_l(y), \ee
where $m=|m|$, and
\be P_l = \frac{1}{2^ll!}\frac{d^l}{dy^l}(y^2-1)^l. \ee
The solution is the well known spherical harmonic function
\be Y_{lm} = N_{lm}P_l^m(y)e^{im\varphi}, \ee
with normalization constant
\be N_{lm}= \sqrt{\frac{(l-m)!(2l+1)}{(l+m)!4\pi}}. \ee
This solution corresponds to the case of an ellipsoid degenerated into
a sphere, and we can put $y=\cos \theta$ for identification with
the angular spherical coordinates $(\theta, \varphi)$.
Equation in $x$ corresponds to the radial part of the
well known solution expressed in terms of Laguerre polynomials.

\subsection{The particular case $q=0$}\l{q0}

In the particular case of zero charge, $q=0$,
we have from Eqs.(\ref{f}) and (\ref{g})
\be\l{f3}
\dx\left((x^2-1)\dx f\right)
-\left(\lm - c^2 x^2 + \frac{m^2}{x^2-1}\right)f=0,
\ee
\be\l{g3}
\dy\left((1-y^2)\dy g\right)
+\left(\lm - c^2 y^2 - \frac{m^2}{1-y^2}\right)g=0.
\ee
One can see that these equations originate straightforwardly also from 
the standard wave equation
$\nabla^2\psi+k^2\psi=0$, in the spheroidal coordinates $(x,y,\varphi)$.
Recurrence relations (\ref{fn}) and (\ref{gn}) then become
\be\l{fn3}
c^2f_{n-2}-(\lm-(m+n)(m+n+1))f_n - (n+1)(n+2)f_{n+2}=0
\ee
and
\be\l{gn3}
c^2g_{n-2}-(\lm-(m+n)(m+n+1))g_n - (n+1)(n+2)g_{n+2}=0,
\ee
which are equivalent to each other.

A general solution for $f$ is given by linear combinations
of radial spheroidal functions $R_{mn}^{(p)}(c,x)$
of first, $p=1$, and second, $p=2$,  kind \cite{Abramovits},
\be\label{Rmn}
R_{mn}^{(p)}(c,x)=\left\{\sum_{r=0,1}^{\infty '}{(2m+r)!\over r!}d_r^{mn}
\right\}^{-1}\left({x^2-1\over x^2} \right)^{m/2}\times
\ee
$$
\times \sum_{r=0,1}^{\infty '} {i^{r+m-n}} {(2m+r)!\over r!}d_r^{mn}
Z_{m+r}^{(p)}(cx),
$$
where,
\be
Z_{n}^{(1)}(z)= \sqrt{\pi\over 2z}J_{n+1/2}(z),
\ee
\be
Z_{n}^{(2)}(z)= \sqrt{\pi\over 2z}Y_{n+1/2}(z),
\ee
and $J_{n+1/2}(z)$ and $Y_{n+1/2}(z)$  are Bessel functions
of first and second kind, respectively.
The sum in (\ref{Rmn}) is made over either even or odd
values of $r$ depending on the parity of $n-m$.
Asymptotics of $R_{mn}^{(1)}(c,x)$ and $R_{mn}^{(2)}(c,x)$ are
\be
R_{mn}^{(1)}(c,x )\stackrel{cx\to\infty}{\longrightarrow}{1\over cx}
\cos\left[ cx -{1\over 2}(n+1)\pi\right],
\ee
\be
R_{mn}^{(2)}(c,x)\stackrel{cx\to\infty}{\longrightarrow}{1\over cx}
\sin\left[ cx -{1\over 2}(n+1)\pi\right].
\ee
Particularly, to have well defined limit at $x=0$ we should
use only spheroidal function of first kind, $R^{(1)}_{mn}(c, x)$,
because Bessel function of second kind, $Y_n(z)$, has
logarithmic divergence at $z=0$.

General solution for $g$ is given by linear combination of
angular spheroidal functions of first and second kind \cite{Abramovits},
\be\l{Smn1}
S_{mn}^{(1)}(c,y)=\sum_{r=0,1}^{\infty '}d_r^{mn}(c)P_{m+r}^m(y),
\ee
\be\l{Smn2}
S_{mn}^{(2)}(c,y)=\sum_{r=-\infty }^{\infty'}d_r^{mn}(c)Q_{m+r}^m(y),
\ee
where $P_n^m(y)$ and $Q_n^m(y)$  are the associated
Legendre polynomials of first and second kind, respectively.

Expressions for radial and angular spheroidal
functions, and corresponding eigenvalues $\lm$,
for particular values of $m$ and $n$, are presented in
Ref. \cite{Abramovits}.

Coefficients $d_k^{mn}(c)$ are calculated with the help of the
following recurrence relation:
\be\l{drecurrence}
\alpha_kd_{k+2} + (\beta_k-\lm_{mn})d_k +\gamma_kd_{k-2}=0,
\ee
where
\be
\alpha_k={(2m+k+2)(2m+k+1)c^2\over (2m+2k+3)(2m+2k+5)},
\ee
\be
\beta_k=(m+k)(m+k+1)+{2(m+k)(m+k+1)-2m^2-1\over (2m+2k-1)(2m+2k+3)}c^2,
\ee
\be
\gamma_k={k(k-1)c^2\over (2m+2k-3)(2m+2k-1)}.
\ee
The calculation is made by the following procedure.
First, one calculates $N_r^m$,
\be
N^m_{r+2}=\gamma_r^m -\lm_{mn}- {\beta_r^m\over N_r^m} \quad (r\geq 2),
\ee
\be
N^m_{2}=\gamma_0^m -\lm_{mn};\quad N^m_{3}=\gamma_1^m -\lm_{mn},
\ee
\be
\gamma_r^m=(m+r)(m+r+1) + {1\over 2}c^2\left[1- {4m^2-1\over
(2m+2r-1)(2m+2r+3)} \right] \quad (r\geq 0).
\ee
Second, one calculates the fractions $d_0/d_{2r}$ and $d_1/d_{2p+1}$
with the use of
\be
{d_0\over d_{2r}} = {d_0\over d_{2}}{d_2\over d_{4}}\cdots {d_{2r-2}
\over d_{2r}},
\ee
\be
{d_1\over d_{2p+1}} = {d_1\over d_{3}}{d_3\over d_{5}}\cdots {d_{2p-1}
\over d_{2p+1}},
\ee
and
\be
N_{r}^m= {(2m+r)(2m+r-1)c^2\over (2m+2r-1)(2m+2r+1)}{d_r\over d_{r-2}}
\ee
The coefficients $d_0$, for even $r$, and $d_1$, for odd $r$,
are determined via the normalization of the solution.

\subsection{The general case}

In this Section, we consider the general solution
of our basic equations (\ref{f}) and (\ref{g}).
To have more general set up, we consider the case of different
charges of nuclei, $Z_1$ and $Z_2$. This leads to appearance of
additional linear in $y$ term in Eq.(\ref{g}), so that both
the ordinary differential equations become of similar structure.
Also, we restrict consideration by analyzing discrete
spectrum, i.e. we assume that the energy $E<0$.

Let us denote
\be\l{pab}
p={R\over 2}\sqrt{-2E},\quad a=R(Z_2+Z_1),\quad b=R(Z_2-Z_1).
\ee
Then, Eqs.(\ref{f}) and (\ref{g}), for the general case
of different charges of nuclei, can be written as
$$
\dx\left((x^2-1)\dx f_{mk}(p,a;x)\right)
$$
\be\l{fp}
+\left(-\lm_{mk}^{(x)}-p^2(x^2-1)+ax-\frac{m^2}{x^2-1}\right)
f_{mk}(p,a;x)=0,
\ee
$$
\dy\left((1-y^2)\dy g_{mq}(p,b;y)\right)
$$
\be\l{gp}
+\left(\lm^{(y)}_{mq}-p^2(1- y^2)+by-\frac{m^2}{1-y^2}\right)
g_{mq}(p,b;y)=0,
\ee
where we assume that the solutions obey
\be
|f_{mk}(p,a; 1)|<\infty , \quad
\lim_{x\to\infty } f_{mk}(p,a; x)=0, \quad
|g_{mq}(p,b;\pm 1)|<\infty.
\ee
The eigenvalues $\lm$ in Eqs.(\ref{fp}) and (\ref{gp}) should be
equal to each other,
\be\l{ll}
\lm_{mk}^{(x)}(p,a)=\lm_{mq}^{(y)}(p,b).
\ee
The general solution $\psi(x,y,\varphi)$ of Eq.(\ref{eqH2plus2})
is represented in the following factorized form:
\be\l{npsi}
\psi_{kqm}(x,y,\varphi ;R)=N_{kqm}(p,a,b)f_{mk}(p,a;x)g_{mq}(p,b;y)
{\exp(\pm im\varphi)\over \sqrt{2\pi}}.
\ee
The normalization coefficients $N_{kqm}(p,a,b)$ in Eq.(\ref{npsi})
are represented with the help of derivatives of the eigenvalues,
$\lm_{mk}^{(x)}(p,a)$ and $\lm_{mq}^{(y)}(p,b)$,
namely,
\be\l{N}
N^2_{kqm}(p,a,b)={16p\over R^3}\left[{\partial\lm_{mq}^{(y)}(p,b)
\over\partial p}-{\partial \lm_{mk}^{(x)}(p,a)
\over\partial p}\right]^{-1}.
\ee
For a given indices $k$, $q$, $m$, and fixed values of $Z_1$, $Z_2$, and
$R$, the discrete energy spectrum $E$ can be determined from Eq.(\ref{ll}).
This equation has unique solution, $p=p_{kqm}(a,b)$.
Then, by solving the relation stemming from (\ref{pab})
\be
p_{kqm}(R(Z_2+Z_1), R(Z_2-Z_1))={R\over 2}\sqrt{-2E}
\ee
in respect to $E$, we can find the discrete spectrum of energy,
\be\l{E}
E_j(R)=E_{kqm}(R,Z_1,Z_2).
\ee
Number of zeroes, $k$, $q$, and $m$, of the functions
$g(y)$, $f(x)$, and $\exp{\pm im\varphi}$
are the angular, radial and azimuthal quantum numbers, respectively.
However, instead of $k$, $q$, and $m$ one can use their linear combinations,
namely, $N=k+q+m+1$ is main quantum number and
$l=q+m$ is orbital quantum number.

To construct the general solution $u(z)$, which is called
{\it Coloumb spheroidal function} \cite{Komarov} ({\sc csf}), in terms of
{\it angular} {\sc csf} $g(y)$ and {\it radial} {\sc csf} $f(x)$,
let us, again, use the form which accounts for singularities at the points
$z=\pm 1$ and $z=\infty$,
\be\l{u}
u(z)=(1-z^2)^{m/2}\exp [-p(1\pm z)]v(z).
\ee
Then, we represent $v(z)$ as an expansion,
\be\l{v}
v(z)=\sum_{s=0}^{\infty} a_s(p,b,\lm)w_s(z),
\ee
in some set of basis functions $w_s(z)$.

Now, the complexity of the recurrence relations depends on the basis.
In the preceding sections, where the particular cases, $R=0$ and $q=0$,
have been considered, we used a power series representation.
One can try other forms of the representation as well.
For a good choice of the basis functions $w_s(z)$,
we can obtain {\it three-term} recurrence relation of the form
\be\l{req}
\alpha_s a_{s+1}-\beta_s a_s +\gamma_s a_{s-1}=0,
\ee
where $\alpha_s$, $\beta_s$, and $\gamma_s$ are some polynomials
in $p$, $b$, and $\lm$. Then, using the tridiagonal matrix $\hat A$
consisting of the coefficients $\alpha_s$, $\beta_s$, and $\gamma_s$
entering Eq.(\ref{req}), we can write down the equation to find out
eigenvalues $\lm_{mk}^{(x)}(p,a)$
and $\lm_{mq}^{(y)}(p,b)$. Namely,
\be\l{detA}
{\rm det}\ \hat A=F(p,b,\lm)=0.
\ee
The matrix $\hat A$ has a tridiagonal form. This leads directly
to one-to-one correspondence between ${\rm det}\ \hat A$ and
the {\it infinite chain fraction},
\be\l{cf}
F(p,b,\lm )=
\beta_0 - \frac{\alpha_0\gamma_1}{\displaystyle
\beta_1-\frac{\alpha_1\gamma_2}{\displaystyle
\beta_2-\ldots\frac{\alpha_N\gamma_{N+1}}{\beta_{N+1}-\ldots}{\displaystyle
}}}
=\beta_0 - {\alpha_0\gamma_1\over \beta_1-}
{\alpha_1\gamma_2\over \beta_2-}\ldots
\simeq\frac{Q_N}{P_N}.
\ee
In numerical computations, this relation allows one to find out
eigenvalues $\lm$ in an easier way due to simpler algorithm
provided by the chain fraction. Consequently, one can compute
the energy and coefficients $a_s$ of the expansion of eigenfunctions
$g(y)$ and $f(x)$ by using the chain fraction.

The result of this approach in constructing of the solutions
depends on the convergence of the chain fraction.
Analysis of the convergence can be made from a general point of view.
Sufficient conditions of the convergence of
the chain (\ref{cf}), and of the expansion (\ref{v}),
are the following two relations:
\be
\left|{\alpha_{s-1}\gamma_s\over \beta_{s-1}\beta_s}\right| <{1\over 4},
\quad
{a_{s+1}\over a_s}_{|s\to\infty}\sim {\beta_s\over 2\alpha_s}
\left[1-\left(1-4{\alpha_s\gamma_s\over\beta_s^2}\right)^{1/2}\right].
\ee
Further analysis of the convergence depends on specific choice of the
basis functions $u_s(z)$.

(i) Series expansion, $v_s(z)=z^s$. In this case,
the radius, $Z_v$, of convergence is
\be
Z_v=\lim_{s\to\infty}\left|{a_s\over a_{s+1}}\right|.
\ee
Particularly, when $a_{s+1}/a_s \to 0$ at $s\to \infty$ the series
(\ref{v}) converges at any $z$.

(ii) For the choice of basis function $v_s(z)$ in the form of
orthogonal polynomials, the sufficient condition for convergence of
Fourier series (\ref{v}) is
\be
\left|{a_s\over a_{s+1}}\right|_{s\to\infty}\leq 1-{1\over s}.
\ee
Below, we consider separately angular and radial {\sc csf} entering the
general solution.

\subsubsection{The angular Coloumb spheroidal function}

For the angular Coloumb spheroidal function ({\sc acsf}), it is natural
to choose the basis functions $v_s(y)$ in the form of
associated Legendre polynomials, $P_{s+m}^m(y)$.
Indeed, they form complete system in the region $y\in [-1, 1]$, and
reproduce {\sc acsf} at $p=b=0$ (see Sec.~\ref{R0}).
Inserting of the expansion
\be
g_{mq}(p,b;y)=\sum_{s=0}^{\infty}c_s P_{s+m}^m(y)
\ee
into Eq.(\ref{gp}) entails five-term recurrence relation.
However, this relation, which is sometimes used, 
is not so suitable as the three-term relation.
This is because the determinant of the corresponding pentadiagonal matrix
can not be represented as a chain fraction.
Nevertheless, in the case $b=0$, i.e. for $Z_1=Z_2$,
this five-terms recurrence relation is reduced to two
three-terms recurrence relations, separately for even
($c_{-2}=0$, $c_0=1$)  and odd
($c_{-1}=0$, $c_1=1$) solutions of Eq.(\ref{gp}) presented in previous
Section.

For the general case $b\neq 0$, the expansions of $g(p,b;y)$,
handling singularities at the points $y=\pm 1$ and $y=\infty$,
respectively, as considered by Baber and Hasse \cite{Baber}, are
\be\l{pl}
g_{mq}(p,b;y)=\exp[-p(1+y)]\sum_{s=0}^{\infty}c_s P_{s+m}^m(y),
\ee
\be\l{min}
g_{mq}(p,b;y)=\exp[-p(1-y)]\sum_{s=0}^{\infty}c_s' P_{s+m}^m(y),
\ee
These expansions yield three-terms recurrence relation,
\be\l{rc}
\rho_sc_{s+1}-\kappa_s c_s + \delta_s c_{s-1}=0, \quad c_{-1}=0,
\ee
where the coefficients for the case of expansion (\ref{pl}) have the
following form:
$$
\rho_s={(s+2m+1)[b-2p(s+m+1)]\over 2(s+m)+3},
$$
\be\l{coef}
\kappa_s = (s+m)(s+m+1)-\lm,
\ee
$$
\delta_s={s[b+2p(s+m)]\over 2(s+m)-1}.
$$
To estimate convergence of these expansions, one can use the
above made estimation of the convergence, with the following
replacements: $\alpha_s\to \rho_s$, $\beta_s\to \kappa_s$,
$\gamma_s\to \delta_s$, and $a_s\to c_s$. For the expansion
(\ref{+}) we have
\be
\left|{\rho_{s-1}\delta_s\over \kappa_{s-1}\kappa_s}
\right|_{s\to\infty}\sim \left({p\over s}\right)^2,
\ee
i.e., at $p>1$, convergence takes place only at $s>2p$.
We should take into account this condition when choosing
{\it minimal} number of terms in the chain fraction (\ref{cf'})
which is sufficient to calculate $\lm$, to a required
accuracy.

The recurrence relation for the coefficients $c_s'$ of the expansion
(\ref{min}) differs from that of Eq.(\ref{rc}) by the replacement
$p\to -p$ in formulas (\ref{coef}). Clearly, this replacement
does not change the form of the chain fraction,
\be\l{cf'}
F^{(y)}(p,b,\lm )=\kappa_0-{\rho_0\delta_1\over \kappa_1-}
{\rho_1\delta_2\over\kappa_2-}\cdots
\ee
So, in both the cases, (\ref{pl}) and (\ref{min}), the eigenvalues
$\lm$ can be found from one and the same equation,
\be\l{Fl}
F^{(y)}(p,b,\lm)=0.
\ee
In practical calculations with the help of this algorithm,
the infinite chain fraction (\ref{cf}) is, of course, replaced
by the {\it finite} one, $F^{(y)}_{N+1}(p,b,\lm )$, in which
one retains a sufficiently big number $N$ of terms. Typically,
$N>10$ provides very good accuracy. 
So, the eigenvalues are computed as the roots of the polynomial
$Q_{N+1}(p,b,\lm)$ of degree $N+1$, namely,
\be\l{FlN}
F^{(y)}_{N+1}(p,b,\lm )=
{Q_{N+1}(p,b,\lm )\over P_{N+1}(p,b,\lm )}.
\ee
Such a representation allows one to exclude singularities,
associated to zeroes of the polynomial $P_{N+1}(p,b,\lm)$,
from Eq.(\ref{Fl}).
Further, from the definitions (\ref{cf}) and (\ref{FlN})
we obtain the following recurrence relation for the polynomial
$Q_{k}(p,b,\lm)$:
\be\l{Qk}
Q_{k+1}=Q_k \bar\kappa_{N-k}-Q_{k-1}\bar\rho_{N-k}\bar\delta_{N-k+1},
\quad Q_{-1}=0,\quad Q_0=1,
\ee
with the use of which one can find $Q_{N+1}$.
Here, the coefficients
$\bar\kappa_{s}$, $\bar\rho_{s}$, and $\bar\delta_{s}$
differ from that of Eq.(\ref{coef}) by the factor $(1+\kappa_s^2)^{-1/2}$.
This factor does not change the recurrence relation (\ref{rc}).
However, it makes possible to avoid accumulating of big numbers
at intermediate computations.
Indeed, from Eq.(\ref{coef}) for $\kappa_s$ it follows that the
leading coefficients of the polynomials $Q_k$ would behave as $k^{4k}$,
for example, for $k=4$ we would have $4^{16}$,
if we would not made the above mentioned
renormalization of the coefficients
$\rho_{s}$, $\kappa_{s}$, and $\delta_{s}$.
The eigenvalue is found as an appropriate root of the polynomial
$Q_{N+1}(p,b,\lm^{(y)})$. Clearly, for big $N$, there is no way to
represent in general the roots of $Q_{N+1}$ analytically so
one is forced to use numerical computations.

In the numerical computations, to pick up the appropriate eigenvalue
$\lm_{mq}^{(y)}(p,b)$ among $N+1$ roots of the
polynomial $Q_{N+1}$ it is necessary to choose some
starting value of $\lm$.
For example, one can put the starting value at the point
$p=b=0$, where $\lm_{mq}^{(y)}(0,0)=(q+m)(q+m+1)$.
The first step is to increase discretely $p \to p+\Delta p$ and
$b\to b+\Delta b$ beginning from the starting point $p=b=0$,
at fixed values of $m$ and $q$, and the second step is to 
find $\lm_{mq}^{(y)}(p+\Delta p,b+\Delta b)$
with the help of Eq.(\ref{Fl}). Repeating these steps one can find
$\lm_{mq}^{(y)}$ numerically as a function of $p$ and $b$
in some interval of interest.

Also, asymptotics of $\lm$ which will be studied in Sec.~\ref{Asymptotics}
are of much help here to choose the appropriate root.
For example, for $b=0$ and $N=5$ we obtain numerically from
the determinant of the tridiagonal matrix consisting of
the coefficients defined by Eq.(\ref{coef}), with $-\kappa_s$ on
the main digonal, and $\rho_s$ and $\delta_s$ on the upper and
lower adjacent diagonals respectively, the polynomial,
\be
\mbox{det}\hat A = 0.003\lm^6 - 0.2\lm^5+(0.2p^2+5.5)\lm^4
-(6.3p^2+56)\lm^3+(1.5p^4+66p^2+231)\lm^2
\ee
$$-(19p^4+226p^2+277)\lm + p^6+44p^4+186p^2.$$
Only one of its six roots has asymptotics,
\be
{\lm}_{|p\to0} = 0.667p^2-0.0148p^4+O(p^5),
\ee
which reproduces, to a good accuracy, the asymptotics (\ref{l00}).
So, this is the desired root to be used in subsequent calculations.
Also, observe the decrease of the numerical coefficients
at higher degrees of $\lm$ which control the convergence.

Note that, at $p\gg 1$, the {\sc acsf} is concentrated
around the points $y=\pm 1$ so that expansion
(\ref{pl}) converges slowly. In this case one uses another,
more appropriate, expansions,
\be\l{+}
g_{mq}(p,b;y)=(1-y^2)^{m/2}\exp[-p(1+y)]\sum_{s=0}^{\infty}c_s (1+y)^s,
\ee
\be\l{-}
g_{mq}(p,b;y)=(1-y^2)^{m/2}\exp[-p(1-y)]\sum_{s=0}^{\infty}c_s'(1-y)^s.
\ee
Evidently, expansion (\ref{+}) converges faster in the region
$[-1,0]$ while the expansion (\ref{-}) converges faster in
the region $[0, 1]$. Here, the coefficients $c_s$ of the
expansion (\ref{+}) obey the three-term recurrence sequence (\ref{rc}),
with
$$
\rho_s={2(s+1)(s+m+1)},
$$
\be\l{coeff}
\kappa_s=s(s+1)+(2s+m+1)(2p+m)+b-\lm,
\ee
$$
\delta_s=b+2p(s+m).
$$
It is remarkable to note that expansions (\ref{+}) and (\ref{-})
converge at any $y$, and the corresponding chain fractions
(\ref{cf}) converge at any $p$ since
\be
{c_{s+1}\over c_s}_{|s\to\infty}
\sim {2p\over s}, \quad
\left|{\rho_{s-1}\delta_s\over \kappa_{s-1}}\kappa_s\right|_{s\to\infty}
\sim {4p\over s}.
\ee
Similarly, the coefficients $c_s'$ obey the same relation, with
the replacement $b\to -b$ in Eq.(\ref{coeff}).

In practical calculations, one can use a combination of
expansions (\ref{pl}) and (\ref{+}). Namely, the procedure is:
from expansion (\ref{pl}) one finds eigenvalues
while the eigenfunctions are calculated from to Eq.(\ref{+}).
Of course, both solutions (\ref{+}) and (\ref{-})
should be sewed, for example, at the point $y=0$,
because the recurrence relations do not determine, in this case,
a general normalization of the coefficients $c_s$ and $c_s'$.
Particularly, the sewing condition, which defines the normalization
of $c_s$ and $c_s'$, has the form
\be
\sum_{s=0}c_s=\sum_{s=0}c_s'.
\ee

To derive the asymptotics of {\sc acsf} and its eigenvalues we can use
an expansion in Laguerre polynomials,
\be\l{Lag}
g_{mq}(p,b;y)=(1-y^2)^{m/2}\exp[-p(1\pm y)]\sum_{s=0}^{\infty}
c_s L_{s+m}^m(2p(1\pm y)),
\ee
\be
L_n^m(z)={e^{z}z^{-m}\over n!}{d^n\over dz^n}(e^{-z} z^{n+m}).
\ee
The insertion of this expansion into Eq.(\ref{gp}) and the use of
the differential equation for Laguerre polynomials,
\be
z{d^2\over dz^2}L_n^m(z)+(1-z+m){d\over dz}L_n^m(z)+nL_n^m(z)=0
\ee
yield recurrence relation (\ref{rc}).
For the case of positive sign in Eq.(\ref{Lag}), we should put
$$
\rho_s=-(s+m+1)\left(s+1+{b\over 2p}\right),
$$
\be\l{coefL}
\kappa_s = -(2s+m+1)\left(s+m+1+{b\over 2p}-2p\right)+(s+m)(m+1)+b-\lm,
\ee
$$
\delta_s=-s\left(s+m+{b\over 2p}\right).
$$

\subsubsection{The radial Coloumb spheroidal function}

The radial Coloumb spheroidal function ({\sc rcsf}) obviously should be
written in a form suitable to handle singularities at the points $x=1$ and
$x=\infty$, namely,
\be\l{RCSF}
f_{mk}(p,a;x)=(1-x^2)^{m/2}\exp[-p(x-1)]f(x).
\ee
So, the equation for $f(x)$ takes the form
\be
(x^2-1)f''(x)+[-2p(x^2-1)+2(m+1)x]f'(x)+[-\lm+m(m+1)+2p\sigma x]
f(x)=0,
\ee
where we have denoted $\sigma = \frac{a}{2p}-(m+1)$.
In the case when the expansion
\be\l{fx}
f(x)=\sum_{s=0}a_s u_s(x)
\ee
implies a three-terms recurrence relation, the eigenvalues
$\lm_{mk}^{(x)}(p,a)$ can be found from the chain fraction equation,
\be\l{Fla}
F^{(x)}(p,a;\lm)=0.
\ee
Also, the expansion which is of practical use has been considered
by Jaffe \cite{Jaffe}. In this case, the expansion series (\ref{fx})
becomes
\be\l{Jaf}
f(x)=(x+1)^{\sigma }\sum_{s=0}a_s \chi^s,
\ee
where $\chi =(x-1)/(x+1)$ is Jaffe's variable.
By inserting (\ref{Jaf}) into the equation for the
function $f(x)$, we get recurrence relation (\ref{req}), where the
coefficients are
$$
\alpha_s=(s+1)(s+m+1),
$$
\be\l{coefJ}
\beta_s = 2s^2+(2s+m+1)(2p-\sigma )-a-m(m+1)+\lm =
\ee
$$
=2s(s+2p-\sigma)-(m+\sigma)(m+1)-2p\sigma+\lm,
$$
$$
\gamma_s=(s-1-\sigma)(s-m-1-\sigma).
$$
Also, for Jaffe series expansion, we have
\be
\left|{\alpha_{s-1}\gamma_s\over \beta_{s-1}\beta_s}
\right|_{s\to\infty}={1\over 4}\left(1-{4p\over s}\right)
+O\left({p^2\over s^2}\right),
\ee
i.e., the chain fraction converges at $p>0$. One can see also that
the Jaffe expansion converges at any $x$.

In addition, the function $f(x)$ can be expanded in associated Laguerre
polynomials,
\be\l{assLag}
f(x)=(x+1)^{\sigma }\sum_{s=0}^{\infty}a_s L_{s+m}^m (\bar x), \quad
\bar x=2p(x-1).
\ee
In this case,the  recurrence relation is of three-terms form, and the
coefficients are
$$
\alpha_s=-(s+m+1)\left[{a\over 2p}-(s+1)\right]=(s+m+1)(s-m-\sigma),
$$
\be\l{cLag}
\beta_s =-(2s+m+1)\left[{a\over 2p}-(s+m+1)\right]+
\ee
$$
+ 2p(2s+m+1)-(s+m)(m+1)-a+\lm,
$$
$$
\gamma_s=-s\left[{a\over 2p}-(s+m)\right]=s(s-1-\sigma).
$$
As to numerical computation of the eigenvalues $\lm^{(x)}$, expansions
(\ref{Jaf}) and (\ref{assLag}) are equivalent because
the chain fraction depends, in fact, only on $\beta_s$ and
$\alpha_s\gamma_{s+1}$. Indeed, by comparing
Eq.(\ref{coefJ}) and Eq.(\ref{cLag}), one can easily see
that in both cases $\beta_s$ and $\alpha_s\gamma_{s+1}$
are the same. Evidently, it then follows that the associated chain 
fractions are equivalent to each other.

However, we should note that
Jaffe's recurrence sequence, in general, is more stable, while
Laguerre expansion (\ref{assLag}) is more suitable to find out the
asymptotics of $f_{mk}(p,a;x)$. 

Also, we note that the associated "radial" polynomials
$Q_{N+1}$, the root $\lm^{(x)}(p,a)$ of which should be found, contain a
much bigger number of terms, in comparison to the "angular" case.
So, practically finding of radial eigenvalues is much harder
than that of angular eigenvalues.

At equal charges of nuclei, $Z_1=Z_2$, the equation for $g$,
and the recurrence relation for $g_k$, are
the same as they are in the particular case $q=0$ considered
in Sec.~\ref{q0}.
A general solution for $g$ is then given by {\sc acsf}
(\ref{Smn1}) and (\ref{Smn2}), with coefficients $d_r^{mn}$
given by recurrence relation (\ref{drecurrence}).

In the reminder of this Section we would like to note that, in general, 
solving the recurrence relations can be made
equivalent to solving associated ordinary differential equations
by making the $z$ transform. In many cases the $z$ transform helps
to solve recurrence relations. Namely, one defines the function
\be
Z(z)=\sum\limits_{s=0}^{\infty} \frac{a_s}{z^n}
\ee
associated to the coefficients $a_s$ entering Eq.(\ref{req}) viewed
as a function of discrete variable $s$.
For $\alpha_s$, $\beta_s$, and $\gamma_s$ given by Eq.(\ref{coefJ})
we obtain from Eq.(\ref{req})
\be
z(z-1)^2Z''
+\left[(1-m)z^2+2(2p-\sigma-1)z+2\sigma+m+1\right]Z'+
\ee
$$
+\left[
(\sigma+m)(m+1)+2p\sigma-\lambda+\frac{(m+\sigma)\sigma}{z}\right]Z = 0.
$$
For the coefficients $c_s$, we define
\be
Y(z)=\sum\limits_{s=0}^{\infty} \frac{c_s}{z^n}.
\ee
and for $\rho_s$, $\kappa_s$, and $\delta_s$ given by Eq.(\ref{coeff})
we obtain from Eq.(\ref{rc})
\be
z(z-1)^2Y''
+\left[(\tau-2)z^2+2(2p-\tau+1)z+\tau\right]Y'+
\ee
$$
+\left[(m(m+1)-m\tau)z- 2p(m+1)+(\tau+m)(m+1)+b-\lambda
-\frac{\tau}{z}\right]Z-
$$
$$
-m(m-\tau+1)c_0z = 0,
$$
where we have denoted
\be
\tau= \frac{b}{2p}+m+1.
\ee
If one has solved these differential equations for $Z(z)$
and $Y(z)$, then, by making the inverse $z$ transform, 
one can find the expansion coefficients $a_s$ and $c_s$
(and thus the general solution of the problem).

\subsection{Asymptotics of {\sc csf} and their eigenvalues}\l{Asymptotics}

To analyze the exact solution, which is of rather complicated
{\it nonclosed} form (infinite series) given in the previous Sections, 
it is much instructive to derive its asymptotics, which can be represented 
in a closed form.
In this Section, we present the asymptotics at large ($R\to\infty$)
and small ($R\to 0$) distances between the nuclei,
with a particular attention paid to the ground state.

\subsubsection{Asymptotics at $R\to\infty$}

For increasing distances $R$ between the nuclei, at fixed  quantum numbers
$k$, $q$, and $m$, we have increasing values of the parameters
$p$, $a$, and $b$,
\be
p=(-2E)^{1/2}R/2\to\infty, \quad
a=(Z_2+Z_1)R\to\infty, \quad
b=(Z_2-Z_1)R\to\pm\infty.
\ee
Let us introduce the notation
\be
\alpha={a\over 2p}={Z_2+Z_1\over\sqrt{-2E}},\quad
\beta={b\over 2p}={Z_2-Z_1\over\sqrt{-2E}}
\ee
and assume that $\alpha \sim 1$ and $\beta \sim 1$.\\

{\bf {\sc acsf} at $R\to\infty$}.\\
Let us consider the asymptotic expansion of {\sc acsf}.
In this case, the equation for the Whittaker function, $M_{\kappa,\mu}(y)$,
builds ansatz around the poles $y=\pm 1$. Here, the solution is constructed
in two overlapping intervals,
${\cal D}_-=[-1, y_1]$ and
${\cal D}_+=[y_2, 1]$, with $y_2<y_1$.
Then, the asymptotics of {\sc acsf}
$g_{mq}(p,2p\beta ;y)$ in the interval ${\cal D}_-$ have the
form
$$
g_{mq}(p,2p\beta ;y)={d_-\over\Gamma(m+1)}\left[{2\Gamma
\left(\kappa+{1+m\over 2} \right)\over\Gamma\left(\kappa+{1-m\over 2}
\right)}\right]^{1/2}\times
$$
\be\l{D-}
\times{M_{\kappa,m/2}\left(2p(1+y)+2(\kappa +\beta)\ln{1-y\over 2}\right)
\over\sqrt{1-y^2}}[1+O(p^{-1})], \quad y\in {\cal D}_-.
\ee
while in the interval ${\cal D}_+$ it is
$$
g_{mq}(p,2p\beta;y)={d_+\over\Gamma (m+1)}\left[{2\Gamma \left(\kappa'+
{1+m\over 2}\right)\over\Gamma\left(\kappa'+{1-m\over 2}\right)}\right]
^{1/2}\times
$$
\be\l{D+}
\times{M_{\kappa',m/2}\left(2p(1-y)+2(\kappa'+\beta)\ln{1+y\over 2}\right)
\over\sqrt{1-y^2}}[1+O(p^{-1})], \quad y\in {\cal D}_+.
\ee
Here, the coefficients $d_-$ and $d_+$ ($d_-^2+d_+^2=1$)
are defined by the relations
\be
d_-=\left|{\sin{\pi}(2\kappa'-m-1)\over \sin{\pi}(2\kappa-m-1) +
\sin{\pi}(2\kappa'-m-1)}  \right|^{1/2} \times
\ee
$$
\times {\rm sgn}\ \left[{-\cos{\pi}(\kappa-(m+1)/2)\over
\sin{\pi}(\kappa'-(m+1)/2)}\right]
$$
\be
d_+=\left|{\sin{\pi}(2\kappa-m-1)\over \sin{\pi}(2\kappa-m-1)+
\sin{\pi}(2\kappa'-m-1)}\right|^{1/2}.
\ee

{\bf {\sc rcsf} at $R\to\infty$}.\\
Now, let us consider the asymptotic expansion of {\sc rcsf}.
The replacements $x\to -y,$ $p\to -p,$ $\alpha\to -\beta$
convert the radial equation around the point
$x=1$ to the angular equation around the point $y=-1$.
Thus, the corresponding asymptotics of {\sc rcsf} are directly  related
to the above found asymptotics of {\sc acsf}.

The {\sc rcsf}, normalized to the first order in $p$, has the form
$$
f_{mk}(p,2p\alpha ;x)={1\over m!}\left[{2(k+
+m)!\over k!(x^2-1)}\right]^{1/2} \times
$$
\be\l{fM}
\times M_{\kappa,m/2}\left(2p(x-1)\right)[1+O(p^{-1})].
\ee
Since the first index of Whittaker function
in Eq.(\ref{fM}) is $\kappa=k+(m+1)/2$,
the function can be expressed in terms of Laguerre polynomials.\\

{\bf Energy at $R\to\infty$}.\\
In the limit $R\to \infty$, the Coloumb two-center problem is
evidently reduced to two separate problems of Coloumb centers, with
the charges $Z_1$ and $Z_2$. Each of the atoms,
$eZ_1$ and $eZ_2$, is characterized by a set
of parabolic quantum numbers,
$[n,n_1,n_2,m]$ and $[n',n'_1,n'_2,m]$,
which are related to each other by the relations
\be
n=n_1+n_2+m+1,\quad n'=n'_1+n'_2+m+1.
\ee
The number $k$ of zeroes of {\sc rcsf} coincides with
the number $n_1$, for the angular functions of the left center, $eZ_1$,
and with the number $n_1'$,  for the angular functions of the right center,
$eZ_2$.

A series expansion in inverse power of $R$ can be obtained in the form
\be\l{Ennn}
E_{[nn_1n_2 m]}(Z_1, Z_2, R)= -{Z_1^2\over 2n^2}-{Z_2\over R}+
{3Z_2n\Delta\over 2R^2Z_1}- {Z_2n^2\over 2R^3Z_1^2}(6\Delta^2-n^2+1)+
\ee
$$
+{Z_2n^3\over 16R^4Z_1^4}[Z_1\Delta (109\Delta^2-39n^2-9m^2+59)
-Z_2n(17n^2-3\Delta^2-9m^2+19)]+
$$
$$
+{\varepsilon_5\over R^5}
+{\varepsilon_6\over R^6}+ O\left({1\over R^7}\right),
$$
where $\Delta =n_1-n_2$, and $\varepsilon_{5,6}$ are defined
via the expressions,
\be
\varepsilon_5=
-{n^3\over 64Z_1^3}[n_Z(1065\Delta^4 -
594n^2\Delta^2 +1230\Delta^2-234m^2\Delta^2+9m^4+
\ee
$$
+33n^4 -18n^2m^2-18m^2+105-138n^2)+
4n_Z^2\Delta (21\Delta^2 -111n^2 +63m^2-189)].
$$
\be
\varepsilon_6=
-{n^4\over 64Z_1^4}[n_Z\Delta(-2727\Delta^4 +
2076n^2\Delta^2 -5544\Delta^2+1056m^2\Delta^2-93m^4-
\ee
$$
-273n^4 +78n^2m^2+450m^2-1533+1470n^2)
  +2n_Z^2(-207\Delta^4 +1044n^2\Delta^2+
$$
$$
+2436\Delta^2-576\Delta^2m^2-42n^2+371
-162m^2+42m^2n^2-89n^4+15m^4)+
$$
$$
+2n_Z^3\Delta (3\Delta^2-69n^2-117-33m^2)],
$$
where $n_Z=nZ_2/Z_1$.

Eq.(\ref{Ennn}) gives the multipole expansion in the
electrostatic energy of the interaction between the atom $eZ_1$
and the far-distant charge $Z_2$ (so called $eZ_1$-terms).

Note that expansion (\ref{Ennn}) can be obtained
by ordinary perturbation techniques as well.
Indeed, the degrees of $Z_1$ display
the orders of the multipole moment of the atom $eZ_1$.

The series of terms corresponding to the other atom, $eZ_2$, 
is obtained from Eq.(\ref{Ennn}) with the use of self-evident
replacements,
$Z_1\leftrightarrow Z_2$, $n\to n'$, $\Delta\to \Delta'$, and
$n_2\to n_2'$.

Finally, the energy of the ground state $1s\sigma_g$ of the molecular ion,
for which $Z_1=Z_2=1$ (equal charges of nuclei), can be written, 
to a high accuracy, as
\be\l{E00}
E_{1000}(1,1,R)=-{1\over 2} - {9\over 4R^4} - {15\over 2R^6}-
{213\over 4R^7}-
{7755\over 64R^8}- {1733\over 2R^9}- {86049\over 16R^{10}}
-O\left({1\over R^{11}}\right).
\ee

\subsubsection{Asymptotics at $R\to 0$}

{\bf Energy at $R\to 0$}.\\
In the case of positive total charge, $Z=Z_1+Z_2>0$, and at $R\to 0$,
we can use perturbative approach to $Z_1eZ_2$ problem,
without using a separation of variables.
Namely, the Hamiltonian of the system $Z_1eZ_2$ is represented as the sum
\be
\hat H=\hat H^{UA}+\hat W= {\hat P^2\over 2m}-{Z_1\over r_1}-{Z_2\over r_2}.
\ee
The operator $\hat H^{UA}$ is usually chosen as the Hamilton operator
of the so called {\it united atom},
\be
\hat H^{UA}= {\hat P^2\over 2m}-{Z\over r_c},
\ee
which is placed on the $z$-axis at the point $z=z_0$,
\be
z_0=\left(-{1\over 2}+{Z_2\over Z}\right)R=
\left({1\over 2}+{Z_1\over Z}\right)R .  \l{z0}
\ee
The point $(0,0,z_0)$ is called {\it center of charges}
due to the fact that it lies at the distances
\be
R_1={Z_2\over Z}R \quad \mbox{ and } \quad R_2={Z_1\over Z}R,
\ee
from the left and right atoms, respectively.

We choose a spherical coordinate system,
$(r_c, \vartheta_c, \varphi)$,
with the origin at point $(0,0,z_0)$, and the angle $\vartheta_c$
measured from $z$-axis. Then, the eigenstates $\psi^{UA}_{Nlm}$
of the operator $\hat H^{UA}$ are
\be
\psi^{UA}_{Nlm}({\vec r_c})=R_{Nl}(r_c)Y_l^m(\vartheta_c, \varphi),
\ee
while the eigenvalues are given by
\be 
E_{Nlm}^{UA}=- {Z^2\over 2N^2}. 
\ee
The matrix $W_{Nlm}^{Nl'm'}$ of the perturbation operator $\hat W$
is diagonal on the functions $\psi^{UA}_{Nlm}({\vec r_c})$ of the atom
if $z_0$ is defined by Eq.(\ref{z0}).
Below, the first two terms of the expansion of energy
in powers of $R$ are given,
$$E_{Nlm}(Z_1,Z_2,R)=$$
\be\l{Bete}
-{Z^2\over 2N^2}-{2Z_1Z_2[l(l+1)-3m^2]
\over N^3l(l+1)(2l-1)(2l+1)(2l+3)}(ZR)^2 + O((ZR)^3).
\ee
For the ground state of the $Z_1eZ_1$ system, with
equal values of the charges, one can find the following
expression for the energy, up to the second order of perturbation:
$$E^{(2)}_{000}(Z_1,Z_1,R)=$$
\be\l{PT2}
{Z^2}\left[-{1\over 2}+{1\over 6}(ZR)^2-
{1\over 6}(ZR)^3+ {43\over 2160}(ZR)^4 - {1\over 36}(ZR)^5\ln{ZR}+ \cdots
\right].
\ee

{\bf {\sc csf} at $R \to 0$}.\\
For fixed quantum numbers we have, at $R\to 0$,
\be
p=(-2E)^{1/2}R/2\to 0, \quad
a=(Z_2+Z_1)R\to 0,  \quad
b=(Z_2-Z_1)R\to 0.
\ee
Let us denote
\be
\alpha ={a\over 2p}={Z_2+Z_1\over\sqrt{-2E}}=\sigma +m+1 ,\quad
\beta ={b\over 2p}={Z_2-Z_1\over\sqrt{-2E}}.
\ee
In this notation, the energy is $E=- Z^2/(2\alpha)^2$.
Let us consider asymptotics of {\sc csf} of the ground state of the
molecular ion. \\

{\bf {\sc acsf} at $R\to 0$}.\\
The power series expansion of {\sc acsf} $g_{00}(p,2p\beta ;y)$
in small parameter $p$ can be obtained by
expanding it in the Legendre polynomials.
For the eigenvalue $\lm_{00}^{(y)}(p,2p\beta)$, we then get
\be\l{l00}
\lm_{00}^{(y)}(p,2p\beta)=(1-\beta^2)\left[ {2\over 3}p^2-
{2\over 135}p^4(1+11\beta^2)+O(p^6) \right].
\ee

{\bf {\sc rcsf} at $R\to 0$}.\\
To expand {\sc rcsf} $f_{00}(p,2p(1+\sigma);x)$, at
$p\to 0$, $\sigma =O(p^2)$, we use Jaffe's expansion,
\be\l{f00}
f_{00}(p,2p(1+\sigma);x)=\exp{(-px)}(1+x)^{\sigma}\sum_{s=0}^{\infty}
a_s\chi^s, \quad
\chi ={x-1\over x+1},
\ee
where $a_s$'s obey three-term recurrence relation
with the coefficients (\ref{coefJ}).

For the eigenvalue $\lm_{00}^{(x)}(p,2p(1+\sigma ))$, we get
\be\l{l00x}
\lm_{00}^{(x)}(p,2p(1+\sigma))=\sigma(1+2p)+\sigma^2
(1+4p\ln{4p\gamma})+o(p^5).
\ee
{\sc rcsf} of the ground state of the molecular ion,
$Z_1=Z_2=1$, can be presented as
\be
f_{00}(p, 2p(1+\sigma); x)=\exp{(-px)}(1+x)^{\sigma}
[1+\sigma^2{\rm Li}_2(\chi ) +o(p^4)],
\l{fLi}
\ee
where $\chi =(x-1)/(x+1)$  is Jaffe's variable and
${\rm Li}_2(\chi)$ is dilogarithm function,
\be
{\rm Li}_2(\chi)=\sum_{n=1}^{\infty}{{\chi }^n\over n^2}=
-\int\limits_0^{\chi}\ln{(1-\xi )}{d\xi\over\xi}.
\ee
The ground state energy is defined as a function of three parameters,
$Z_1$, $Z_2$, and $R$,
\be\l{l=l}
\lm_{00}^{(y)}(p,2p\beta)=\lm_{00}^{(x)}(p,2p(1+\sigma))
\ee
Combining Eqs.(\ref{l00}), (\ref{l00x}) and (\ref{l=l}), we get
series expansion for the ground state energy of the
$Z_1eZ_2$ system in the form
$$E_{000}(Z_1,Z_2,R)=$$
$$
-{1\over2}{Z^2} +{2\over3}Z_1Z_2(ZR)^2-
{2\over3}Z_1Z_2(ZR)^3+{2\over5}Z_1Z_2\left(1-{64Z_1Z_2\over 27Z^2}
\right)(ZR)^4 -
$$
\be\l{ZR}
-{8\over 45}Z_1Z_2\left[{5Z_1Z_2\over Z^2}\ln{(2RZ\gamma)}
+1-{199Z_1Z_2\over 12Z^2}\right](ZR)^5+o((ZR)^5).
\ee
Comparing Eqs.(\ref{PT2}) and (\ref{ZR}) we see that
the terms proportional to $(ZR)^2$ and $(ZR)^3$
coincide. The next order corrections makes a difference;
Eq.(\ref{PT2}) obtained by the second-order perturbation
is of less accuracy.
A practically achieved accuracy of the first-order perturbation (\ref{Bete})
and of Eq.(\ref{ZR}) is the same; at $ZR<0.1$, the discrepancy
is not bigger than 1\%, and becomes sharply smaller with the
increase of the parameter $ZR$.

\subsubsection{Quasiclassical asymptotics}

At $R\to\infty$, for $eZ_1$ solutions we have
\be\l{alx}
\lm_{mk}^{(x)}(p,2p\alpha)=-2p(2\kappa-\alpha)
-\kappa(2\kappa-\alpha-m)+
\ee
$$
+{\kappa\over 2p}(2\kappa^2-3\kappa\alpha+\alpha^2-m^2)+o(p^{-2}),
$$
\be\l{aly}
\lm_{mq}^{(y)}(p,2p\beta)=2p(2\chi+\beta)-\chi(2\chi+2\beta-m)-
\ee
$$
-{\chi\over 2p}(2\chi^2+3\chi\beta+\beta^2-m^2)+o(p^{-2}).
$$
From the equality
$\lm_{mk}^{(x)}=\lm_{mq}^{(y)}$,
we get the expansion for $E_j(R)$
which coincides with the asymptotics (\ref{Ennn}),
up to terms of the order of $R^{-2}$.

In the limit $R\to 0$, the following expansions are justified,
\be\l{0lx}
\lm_{mk}^{(x)}=\left[{a\over 2p}-(k+1/2)\right]^2+O(p^2),
\ee
\be\l{0ly}
\lm_{mq}^{(y)}=(l+1/2)^2+{p^2\over 2}-{b^2\over 8(l+1/2)^2}+O(p^4).
\ee
and we get, by using the equation $\lm^{(x)}=\lm^{(y)}$,
\be\l{0Ej}
E_{Nlm}(R)=-{1\over 2}\left({Z_1+Z_2\over N}\right)^2-R^2
{Z_1Z_2(Z_1+Z_2)^2\over 4N^3(l+1/2)^5}[(l+1/2)^2-3m^2].
\ee
Expression (\ref{0Ej}) for the energy
coincides with the asymptotics (\ref{Bete}), up to $O(l^{-2})$.
Clearly, an accuracy of the quasiclassical Eqs.(\ref{alx})-(\ref{0Ej}) 
becomes higher for a greater number of zeroes of the solutions, $k$ and $q$.
However, even for the ground state, $k=q=m=0$, these equations
give a good approximation for the energy in both limiting cases,
$R\to 0$ and $R\to \infty$. Also, we note that corresponding numerical
calculations showed that for the intermediate values of $R$ the terms
$E_j(R)$ can be determined within the quasiclassical approach with
accuracy of about 5\%, or more \cite{Gerstein}.

\section{Scaling method and binding energy of three-body 
Santilli-Shillady isochemical model $\hat H_2$}\l{Scaling}

To find the ground state energies of $H_2^+$ and $\hat H_2$, we use
computations of the $1s\sigma$ terms  of $H_2^+$ ion
and of $\hat H_2$ based on the above presented exact {\sc csf} solution
by solving the corresponding equations $\lm^{(x)}=\lm^{(y)}$.

 The angular and radial eigenvalues $\lm$ are found as solutions
of the equations containing infinite chain fractions presented 
in previous Sections which should be interrupted and then 
solved numerically, to a required accuracy.

Our primary interest is the study of Santilli-Shillady model
$\hat H_2$ system. However,
we present the results for $H_2^+$ ion as well to check our calculations
and to use them in the scaling method described below.

Also, the reader should keep in mind that we are primarily interested in
ascertaining whether there exist a non-zero value of $R$ for which
a fully stable and point-like isoelectronium permits an exact 
representation of the binding energy of $H_2$ molecule.

It should be noted that the ground state electronic energy 
is obtained as a function of the parameter $R$ due to Eq.(\ref{E}).
By adding to it the internuclear potential energy $1/R$,
we obtain the {\it total} ground state energy of the system,
so that at some value $R=R_{min}$, the total energy $E$ necessarily has
a {\it minimum}, if the system is {\it stable}. This is the way
to determine uniquely the internuclear distance under
an exact representation of the total energy $E$.

To have an independent check of the result for the total ground state 
energy of $\hat H_2$ (with the stable and point-like isoelectronium)
obtaining from the exact solution,
we develop a scaling method based on the original Schr\"odinger equation 
for $H_2^+$ ion like system. Namely, it appears that one is able
to calculate the ground state energy as a function of $R$
for {\it any} $H_2^+$ ion like system with equal charges of nuclei, 
$Z_1=Z_2$, provided that one knows the ground state energy as 
a function of $R$ for the $H_2^+$ ion itself.
It should be pointed out that the scaling method does not depend on 
the obtained solution because it reflects, in fact, the scaling
properties of the Schr\"odinger equation itself.

In addition, we use below Ritz's variational approach to $H_2^+$ like 
systems to find out the {\it approximate} value of the ground state energy
of $\hat H_2$, as well as to check the result provided by the exact solution, 
and to demonstrate the accuracy of the variational approximation.

Our general remark is that in both approaches,
the exact solution and Ritz's variational solution, we use Born-Oppenheimer
approximation (fixed nuclei). Clearly, taking into account the
first order correction, i.e., zero harmonic oscillations of the nuclei
in $H_2$ around their equilibrium positions, we achieve greater 
accuracy.

But we still have a significant inaccuracy in the value of dissociation 
energy due to the fact that $H_2$ system has the lightest 
possible nuclei (two single protons).

To estimate this inaccuracy,
one can invoke Morse's potential customarily used for diatomic molecules.
In particular, the analysis for $H_2$ molecule shows that
the ground state energy of harmonic oscillations of the nuclei
receives 1.4\% correction due to the first anharmonic term.

\subsection{Exact representation of binding energies of $H_2^+$
ion and three-body $\hat H_2$ system}\l{Binding}

The {\sc csf} based computations for $H_2^+$ ion were presented,
for example, by Teller \cite{Teller}, Bates {\it et al.} \cite{Bates},
and Wind \cite{Wind},
and we do not repeat this study here for brevity, while we shall
just describe the procedure and present our final numerical results
in Appendix, Table~\ref{Table2}.

In particular, Teller presented a plot of the resulting function
$E_{1s\sigma}(R)$ which provides a good accuracy, and Wind used
the exact solution to present a table of energy values in seven 
decimal places for distance values of $R$ up to 20 a.u. in steps 
of 0.05 a.u.
However, these results cannot be used directly in our case since
the repulsive potential between the nuclei, $1/R$,
has not been accounted for, and, as the main reason, we have 
the isoelectronium instead of one single electron.

In the Appendix, we present the results obtained
from the {\sc csf} based recurrence relations by numerical
calculations for ordinary $H_2^+$ ion and for $\hat H_2$ system, 
at $M=2m_e$.
These results are presented in Tables~\ref{Table2} and \ref{Table4}.
Tables~\ref{Table3} and \ref{Table5} have been derived
from Tables~\ref{Table2} and ~\ref{Table4}, respectively,
by simple adding internuclear potential energy $1/R$, to obtain 
the {\it total} energy of the system.

In Table~\ref{Table2}, we present the $1s\sigma_g$ electronic term of
$H_2^+$.
In Table~\ref{Table3}, we present the total energy of $H_2^+$.
In Table~\ref{Table4}, we present the $1s\sigma_g$ term of
$\hat H_2$, at the mass $M=2m_e$.
In Table~\ref{Table5}, we present the total energy of $\hat H_2$,
at the mass $M=2m_e$.
Also, in Table~\ref{Table6}, we present the total {\it minimal} energies 
of $\hat H_2$ and {\it optimal} distances $R$, for various values of 
the isoelectronium mass parameter, $M=\eta m_e$.
All the data of these Tables are purely theoretical and, additionally,
we plot them in Figures~1--8,
for the reader convenience. Figures~6 and 8 
give more detailed view on the interval $0.26<\eta<0.34$.

The analysis of the data in Tables~\ref{Table2}--\ref{Table5} is simple.
Namely, one should identify the {\it minimal} value of the energy 
in each Table.
One can use Figures~1--4 
for visual identification of the minima, and then turn to the corresponding 
Tables~\ref{Table2}--\ref{Table5}, to reach a higher numerical accuracy.

Let us consider, as an example, Table~\ref{Table2}.
One can see that the energy minimum for $H_2^+$ is
\be
E_{1s\sigma}=-2.0 \au \mbox{ at } R=0 \au
\ee
Note that $E_{1s\sigma}$ is the electronic energy, i.e. the internuclear
repulsion has not been taken into account here.
Our remark is that this energy value corresponds to $He^+$ ion
due to the fact that two $H_2^+$ nuclei are superimposed and form $He$ 
nucleous at $R=0$.

Let us consider now Table~\ref{Table3}.
In this Table, one can find the line $|2.0 \ |\ -0.602634|$
corresponding to visual minimal value of the energy.
To identify a {\it more precise} value of the minimal energy, one
should use the {\it interpolation} of the data. This gives us
the minimum of the total energy of $H_2^+$,
\be\l{EexactH2p}
E=-0.6026346 \au \mbox{ at }   R=1.9971579 \au
\ee
This theoretical value represents rather accurately the known experimental 
value $E_{exper}[H_2^+] = -0.6017 \au$ \cite{Flugge} for $H_2^+$ ion, 
thus establishing the validity of our {\sc csf} based calculations. 
For completeness, we note 
that the experimental dissociation energy of $H_2^+$ ion is 
$D_{exper}[H_2^+] \simeq 0.0974 \au = 2.65 \eV$, and the internuclear 
distance $R_{exper}[H_2^+]\simeq 2.00 \au=1.0584 \AA$.

Let us now consider Table~\ref{Table4}.
One can see that the energy minimum for $\hat H_2$, at $M=2m_e$, is
\be
E_{1s\sigma}=-16.0 \au \mbox{ at } R=0 \au
\ee
Note that $E_{1s\sigma}$ only yields the isoelectronium's energy,
i.e. the internuclear repulsion has not been taken into account here.
Our remark is that this energy value corresponds to the $He$ atom,
where the two electrons form a stable point-like isoelectronium of mass 
$M=2m_e$.

Let us consider now Table~\ref{Table5} which is of striking interest 
for our study. In this Table, one can find the line
$|0.250 \ |\ -7.61428940411169996|$
corresponding to the visual minimal value of the energy.
To identify a more precise value of the minimal energy, one
should use the interpolation of the data given in this Table.
This gives us the minimum of the total energy of $\hat H_2$,
\be\l{Eexact}
E=-7.617041 \au \mbox{  at  } R=0.258399 \au
\ee
This theoretical value is in quite good agreement
with the preceding theoretical result by Santilli and Shillady
obtained via structurally different variational numerical method,
$E_{var}=-7.61509174$ at $R_{var}=0.2592$ 
(see third column of Table~1 in Ref.\cite{SS}).
It is quite naturally to observe that this variational energy is 
a bit higher (by $0.002 \au$) than the above one obtained from the exact 
solution (as it is expected to be for any variational solution). 

However, this exact theoretical value (\ref{Eexact}) 
does not meet the experimental value
$E_{exper}[H_2] = -1.17 \au$ \cite{Flugge} known for $H_2$ molecule.
Indeed, adopted approximation that the isoelectronium is point-like, 
stable, and has mass $M=2m_e$ leads us to the theoretical value 
(\ref{Eexact}) while the known experimental value, 
$E_{exper}[H_2] = -1.17 \au$, differs much from it.

Essentially the same conclusion is due to numerical program SASLOBE by
Santilli and Shillady \cite{SS},
where Gaussian screened Coloumb potential interaction between
the electrons, rather than the stable point-like isoelectronium 
approximation, has been used to achieve final precise fit of
$E = -1.174474 \au$, with the obtained
bond length $R=1.4011 \au$, at the isoelectronium 
correlation length $r_c=0.01125 \au$ (see Table~1 
in Ref.\cite{SS}). We discuss on this issue in Sec.~\ref{Concluding}.

Our remark is that, due to Table~\ref{Table5},
the experimental value $E=-1.17 \au$ is fitted by the distance 
$R=0.072370 \au$ However, this energy value is not minimal and 
thus can not be ascribed reasonable physical treatment in Table~\ref{Table5}.

Our conclusion from the above analysis is that we have two 
main possibilities to overcome this sharp discrepancy
between our theoretical and the experimental binding energy values
which has place at $M=2m_e$:
\begin{enumerate}
\item Consider unstable isoelectronium, i.e. the four-body 
Santilli-Shillady model of $H_2$ molecule;  
\item Treat the mass $M$ of isoelectronium as a free parameter,
instead of fixing it to $M=2m_e$, assuming thus some defect of mass 
discussed in Introduction,
\end{enumerate}
in order to fit the experimental data on $H_2$ molecule. 

The first possibility will be considered in a subsequent paper
because it needs in application of different technique,
while the second possibility can be studied within the {\it three-body}
Santilli-Shillady model under consideration to which we turn below.

In the next Section, we develop simple formalism allowing one to deal
with the mass and charge of isoelectronium viewed as free parameters,
and arrive at the conclusion (see Table~\ref{Table6}) that
the restricted three-body Santilli-Shillady model of $H_2$ molecule is 
capable to fit the experimental binding energy, 
with the total mass of isoelectronium equal to $M=0.308381m_e$, although 
with the internuclear distance about 19.6\% bigger than the experimental 
value.

\subsection{The scaling method}\l{TheScaling}

In order to relate the characteristics of $H_2^+$ ion like system
to that of thoroughly studied $H_2^+$ ion, we develop scaling method
based on the Schr\"odinger equation.
The neutral $\hat H_2$ system with stable point-like isoelectronium 
is an example of $H_2^+$ ion like system in which we are particularly 
interested here. Below, we develop scaling method for the case
of arbitrary mass and charge of the particle.

 Let us write the Schr\"odinger equation for a particle of 
the rescaled charge 
\be 
e \to -\zeta e, 
\ee
(we turn here from $e=-1$ to $-e=1$ representation), and the rescaled mass 
\be
m_e \to \eta m_e,
\ee
with {\it equal} charges of nuclei, $+eZ_1=+eZ_2=+eZ$,
\be\l{eqS}
\biggl[ -\frac{\hbar^2}{2\eta m_e}\nabla^2_r
-\frac{\zeta Ze^2}{r_a} -\frac{\zeta Ze^2}{r_b}
+\frac{Z^2e^2}{R_{ab}}\biggr]\psi = E\psi.
\ee
where $\eta$ and $\zeta$ are scaling parameters, 
and $R_{ab}$ is distance between the nuclei.
The condition $Z_1=Z_2$ is an essential point to stress here
because owing to which we can successfully develop the scaling method.
We introduce the unit of length,
\be
r_0=\frac{1}{\eta\zeta Z}r_B \equiv \frac{1}{\eta\zeta Z}
\frac{\hbar^2}{m_ee^2},
\ee
where $r_B$ is Bohr's radius.
Dividing Eq.(\ref{eqS}) by $\zeta Z e^2$, and multiplying
it by $r_0$, we get
\be\l{eqS'}
\biggl[ -\frac{\hbar^2}{\eta\zeta Z m_e e^2 }r_0\frac{1}{2}\nabla^2_r
- r_0\frac{1}{r_a} -r_0\frac{1}{r_b}
+r_0\left(\frac{Z}{\zeta}\right)\frac{1}{R_{ab}}\biggr]\psi =
\frac{r_0E}{\zeta Ze^2}\psi.
\ee
We introduce dimensionless entities
$\rho =r/r_0$, $\rho_a =r_a/r_0$, $\rho_b =r_b/r_0$, and
${\cal{R}}=R_{ab}/r_0$. Then, Laplacian in Eq.(\ref{eqS'})
becomes $r_0^2\nabla^2_r= \nabla^2_{\rho}$.
Further, introducing unit of energy,
\be\l{E0}
E_0=\frac{\eta m_e \zeta^2 Z^2 e^4}{\hbar^2} \equiv
\eta\zeta^2 Z^2 \frac{m_ee^4}{\hbar^2},
\ee
we have dimensionless energy $\varepsilon =E/E_0$ so that Eq.(\ref{eqS'})
can be rewritten as
\be\l{eqSdl}
\biggl[ -\frac{1}{2}\nabla^2_{\rho}
- \frac{1}{\rho_a} -\frac{1}{\rho_b}
+\frac{1}{({\zeta\over Z}{\cal{R}})}\biggr]\psi
=\varepsilon\psi.
\ee
Note that, at $\eta=1$, $\zeta=1$, and $Z=1$,
the constants $r_0(\eta,\zeta,Z)$ and $E_0(\eta,\zeta,Z)$
reproduce {\it ordinary} atomic units,
\be
r_0(1,1,1)=r_B=\frac{\hbar^2}{m_ee^2}, \quad E_0(1,1,1)= 2E_B=
\frac{m_ee^4}{\hbar^2},
\ee
and we recover the case of $H_2^+$ ion.
On the other hand, in terms of dimensionless entities the original 
Schr\"odinger equation for $H_2^+$ ion is
\be\label{H_2+}
\biggl[ -\frac{1}{2}\nabla^2_{\rho}
- \frac{1}{\rho_a} -\frac{1}{\rho_b}
+\frac{1}{R}\biggr]\psi_0
=\varepsilon (R)\psi_0,
\ee
where $R=R_{ab}/r_B$.
Comparison of Eq.(\ref{eqSdl}) and Eq.(\ref{H_2+})
shows that by putting $R=(\zeta /Z){\cal{R}}$ in Eq.(\ref{eqSdl}),
we obtain the equation,
\be\label{lH_2+}
\biggl[ -\frac{1}{2}\nabla^2_{\rho}
-\frac{1}{\rho_a} -\frac{1}{\rho_b}
+\frac{1}{R}\biggr]\psi=\varepsilon (R)\psi,
\ee
which identically coincides with the original Eq.(\ref{H_2+}).
The difference is that Eq.(\ref{lH_2+}) is treated in terms of 
the {\it rescaled} units,
$r_0(\eta,\zeta,Z)$ and $E_0(\eta,\zeta,Z)$, 
instead of the ordinary Bohr's units, $r_B$ and $E_B$.
As the result, we have one and the same form of Schr\"odinger equation
for any $H_2^+$ like system characterized by equal charges of nuclei.
This makes a general ground to calculate some characteristic entity
of any $H_2^+$ like system when one knows its value for $H_2^+$ ion.

Particularly, one can easily derive $R_{ab}$ and $E$ for the system
with arbitrary parameters $\eta$, $\zeta$, and $Z$ from
their values, $R_{ab}[H_2^+]$ and $E[H_2^+]=2E_B-\varepsilon(R)$,
obtained for $H_2^+$ ion (for which  $\eta=1$, $\zeta=1$, and $Z=1$).
Indeed, since for arbitrary $\eta$, $\zeta$, and $Z$
\be
R=\frac{\zeta}{Z}{\cal{R}}=\frac{\zeta}{Z}\frac{R_{ab}}{r_0}=
\frac{\zeta}{Z}\frac{R_{ab}}{r_B}\eta\zeta Z ,
\ee
we can establish the following relationship between the distances 
corresponding to arbitrary $Z\zeta Z$ system and $H_2^+$ ion,
\be\l{R}
R_{ab}=\frac{R[H_2^+]}{\eta\zeta^2}.
\ee
It is remarkable to note that the dependence on $Z$ disappeared in
Eq.(\ref{R}). In the case of isoelectronium of mass $M=2m_e$ and 
charge $-2e$, we have $\eta=2$ and $\zeta =2$, so that
\be
R_{ab}=\frac{R[H_2^+]}{8}.
\ee
Also, the energy $E(R)$ of $Z\zeta Z$ system and energy $\varepsilon(R)$
of $H_2^+$ ion are related to each other according to the equation,
\be
E(R)=\eta\zeta^2 Z^2 \left({m_ee^4\over \hbar^2}\right)\varepsilon(R).
\ee

\subsubsection{The case $M=2m_e$}\l{ThecaseM2}

So, in the case of isoelectronium of mass $M=2m_e$ and 
charge $-2e$, we get
\be\l{Eabmin}
E(R_{ab})= 8\varepsilon (R).
\ee
Note however that the factor $\zeta/Z=2$ arised due to
$R=(\zeta /Z){\cal{R}}$
is hidden here so that in order to calculate the values of
$E(R_{ab})$ and $R_{ab}$ from $\varepsilon(R)$ and $R$
respectively one should multiply $\varepsilon$ by 8 and $R$ by $1/4$.

As the result, in accordance with the scaling method the points can be
calculated due to the following rule:
\be\l{scal}
(R,\ E) \to (R, \ E+1/R) \to (R/4, \ 8E) \to (R/4, \ 8E+4/R),
\ee
for Tables \ref{Table2} $\to$ \ref{Table3} $\to$
\ref{Table4} $\to$ \ref{Table5}.
One can easily check numerically that these properties indeed hold true
for the presented Tables.
Thus, the scaling method can be used {\it instead of} the independent
numerical calculations for $\hat H_2$ system if one has the data for
$H_2^+$ ion.

It is highly important to note here that the energy {\it minimum} 
in Table~\ref{Table3} is {\it not} rescaled to the energy {\it minimum} 
in Table~\ref{Table5} due to the absence of energy scaling between 
these Tables; see Eq.(\ref{scal}), from which one can observe 
that $(8E+4/R)$ can not be expressed as $n(E+1/R)$, where $n$
is a number. So one needs to identify minimum in Table~\ref{Table5} 
independently (after calculating all the points), 
rather than direct rescale the minimum from 
Table~\ref{Table3} to try to get minimum for Table~\ref{Table5}.

\subsubsection{The case $M=\eta m_e$}\l{ThecaseMeta}

For a more general case of isoelectronium mass, $M=\eta m_e$,
and charge $-2e$, we should keep the following sequence of calculations:
\be\l{scal2}
(R,\ E) \to (R, \ E+1/R) \to 
(\frac{R}{2\eta}, \ 4\eta E) 
\to (\frac{R}{2\eta}, \ 4\eta E+\frac{2\eta}{R}).
\ee
starting from Table~\ref{Table2} to obtain, at the last step, the 
table of values (similar to Table~\ref{Table5}) from which we should 
extract a {\it minimal} value of the energy and the corresponding 
{\it optimal} distance, at {\it each given} value of mass $\eta$.
The result of the analysis of a big number of such tables
is collected in Table~\ref{Table6}, where 
the interval $0.26<\eta<0.34$ appears to be of interest;
$M=\eta$, in atomic units. Plots of the data of Table~\ref{Table6} are  
presented in Figures~5 and 7 
(Figures~6 and 8 
give more detailed view on the interval of interest) show that
\be
E_{min}(M) \simeq -3.808M, \qquad R_{opt}(M)\simeq \frac{0.517}{M},
\ee
to a good accuracy. Note that $E_{min}(M)$ unboundedly decreases
with the increase of $M$ (there is no local minimum), so we can 
use a fit, instead of the minimization in respect with $M$.
From this Table, we obtain the following final fit of the binding 
energy for the restricted {\it three-body} Santilli-Shillady model of 
$H_2$ molecule:
\be\l{Mfit}
M=0.308381m_e, \quad E=-1.174475 \au , \quad R=1.675828 \au,
\ee
where the mass parameter $M$ of the isoelectronium has been varied
in order to meet the experimental energy 
$E_{exper}[H_2]=-1.174474 \au=-31.9598 \eV$. 
Using this value of mass, $M=0.308381m_e$, we computed the total energy 
as a function of the internuclear distance $R$, and depicted it 
in Figure~9 
to illustrate that $R=1.675828 \au$ indeed corresponds to 
a {\it minimal} value of the energy. Note that the predicted optimal distance 
$R=1.675828 \au = 0.886810 \AA$ appears to be about 19.6\% bigger than the 
conventional experimental value $R_{exper}[H_2]=1.4011 \au = 0.742 \AA$.

This rather big (19.6\%) discrepancy can not be ascribed to 
the Born-Oppenheimer approximation used in this paper since it gives 
relatively small uncertainty in the energy value, even in the case 
of $H_2$ molecule. 
We stress here that in the Born-Oppenheimer approximation, 
the three-body problem (the Schr\"odinger equation) can be given 
exact solution owing to separation of the electronic and nuclear 
degrees of freedom while the full three-body problem 
(accounting for the wave functions of the nuclei, etc.) 
can not be solved exactly. 


In a strict consideration, we should calculate the {\it dissociation} 
energy of $H_2$ molecule, $D=2E_0-E-E^{nucl}$, to make comparison 
to the experimental value, 
$D_{exper}[H_2] \simeq 0.164 \au = 4.45 \eV$ \cite{Flugge}.
Here, $E_0=-0.5 \au =-13.606 \eV$ is the ground state energy of 
separate $H$-atom and $E^{nucl}$ is the energy of zero mode harmonic 
oscillations of the nuclei, with the experimental value 
$E^{nucl}_{exper}[H_2]$ $ \simeq 0.01 \au$ = $0.27 \eV$ \cite{Flugge}. 
One can see that the zero mode energy $E^{nucl}$ (which is taken to be 
$E^{nucl}=0$, in the Born-Oppenheimer approximation) is estimated to be less 
than 1\% of the predicted $E$. The leading anharmonic correction to the 
harmonic oscillation energy is estimated to be 1.4\% of $E^{nucl}$,
i.e. it is of the order of $0.00014 \au=0.004 \eV$, in the case of $H_2$ 
molecule. So, in total the Born-Oppenheimer approximation makes only up to 
1\% uncertainty, which is obviously insufficient to treat the predicted 
$R=1.675828 \au$ as an acceptable value, from the experimental point of view.

Note that, at the given $M$, we can not "fix" $R$ to be equal to the 
desired experimental value $R_{exper}=1.4011 \au$ unless we shift $E$ to some 
{\it nonminimal} value, which is, as such, meaningless. 
Conversely, if we would fit experimental $R_{exper}$ by varying $M$, we were 
obtain $E_{min}$ between 
$-1.52 \au$ and $-1.33 \au$ (see Table~\ref{Table6}), which is much 
deviated from the experimental $E_{exper}$. In other words, the relation 
between $E$ and $R$, governed by the Schr\"odinger equation, 
is such that at some value of $R$ there is a minimum of $E$
so that $R$ is not some kind of free parameter here since the system
tends to minimize its own energy.
In accordance to the exact solution of the model, our single free parameter, 
$M$, can not provide us with the exact fit of {\it both} the experimental 
values, $E_{exper}$ and $R_{exper}$. 

Thus, we arrive at the conclusion that the three-body Santilli-Shillady model 
of $H_2$ molecule yields the result (\ref{Mfit}), which indicates that the 
assumption of stable point-like isoelectronium builds a crude 
{\it approximation} to the general (four-body) Santilli-Shillady model.
This means that we are forced to possess that the isoelectronium is {\it not 
stable point-like} quasi-particle, to meet the experimental data on $H_2$ 
molecule.

\subsection{Variational solution}\l{Variational}

In studying $H_2^+$ ion like systems, one can use Ritz
variational approach to obtain the value of the
ground state energy as well. This approach assumes
analytical calculations, which are easier than that
used in finding the above exact solution but they give {\it approximate}
value of the energy. It is helpful in making simplified
analysis of the system. This can be made
for the general case of isoelectronium total mass, which
eventually undergoes some "defect" while its "bare" total mass is
assumed to be $M=2m_e$.
Ritz variational solution of the $H_2^+$ like problem
yields, of course, similar result for the energy of
$\hat H_2$. Below, we present shortly results of our calculations.
However, we stress that the variational solution is given here just to make 
some support to the exact solution, and to see the typical order of 
the variational approximation.

Using hydrogen ground state wave function and one-parameter Ritz variation,
we obtain the following expression for the energy of $H_2^+$ like system:
\be
E(\rho)=-{1\over 2}{e^2\over a_0} +
{e^2\over a_0}{1\over\rho}{(1+\rho)
e^{-2\rho}+(1-{2\over 3}\rho^2)e^{-\rho}\over
1+ (1+\rho +{1\over 3}\rho^2)e^{-\rho}}, \label{Er}
\ee
where $\rho = R/a_0$ is variational parameter.
For the general case of mass $m$ and charge $q=\zeta e$,
Eq.(\ref{Er}) can be rewritten in the following form:
\be\l{Er1}
E(\rho,\zeta)={Me^4\zeta^2\over \hbar^2}\left(-{1\over 2} + F(\rho )\right),
\ee
where
\be
F(\rho )={1\over\rho}{(1+\rho)e^{-2\rho}+
(1-{2\over 3}\rho^2)e^{-\rho}\over
1+ (1+\rho +{1\over 3}\rho^2)e^{-\rho}}.
\ee
Numerically, the function $F(\rho)$ reaches minimum at the
value $\rho =2.5$, which should be used in the above expression for
$E(\rho,\zeta)$. So, putting $\zeta =1$ we obtain the variational
value of $H_2^+$ ion energy, $E(\rho)=-0.565$. Note, to make
a comparison, that we have the value $E_{exact}=-0.6026$ due to
the exact solution (\ref{EexactH2p}), and the value
$E_{exper}[H_2^+]=-0.6017$ as the experimental value of the energy of
$H_2^+$ ion.
Thus, the optimal distance between the protons in $H_2^+$ ion is
$R_m= a_0 \rho =2.5 \au$, and the obtained variational energy $E$
is slightly higher than both the values $E_{exact}$ and $E_{exper}$,
as it is normally expected to be in the variational approach.
Now, we should replace electron by isoelectronium to describe
the associated $\hat H_2$ model.
Substituting $M=2m_e$ and $\zeta=2$, we see that the r.h.s. of
Eq.(\ref{Er1}) contains overall factor 8, in comparison to the
$H_2^+$ ion case ($M=m_e$ and $\zeta=1$),
\be\l{E2}
\tilde E(\rho)=8|2E_{B}|\left(-{1\over 2} +  F(\rho)\right),
\ee
The function $F(\rho)$ remains the same, and its minimum
is reached again at $\rho =2.5$.
Then, energy of $H_2$ molecule due to Eq.(\ref{E2}) is
$\tilde E(\rho)=-8|2E_{B}|0.565=-4.520 \au$
This value should be compared with the one given by Eq.(\ref{Eexact}).

\medskip
Below, we collect the above mentioned data and results of this
Section in Table~\ref{Table1}.

\begin{table}[ht]
\begin{center}
{\small
\begin{tabular}{|l|l|l|l|}
\hline
                       & {\small $E$, a.u.} & {\small $R$, a.u.} \\
\hline
$H_2^+$ ion, exact theory ($N$=16)          &-0.6026346  &1.9971579\\
$H_2^+$ ion, experiment \cite{Flugge}       &-0.6017     &2.00\\
\hline
3-body $\hat H_2$, $M$=2$m_e$, exact theory ($N$=16) 
                                            &-7.617041   &0.258399\\
3-body $\hat H_2$, $M$=2$m_e$, var. theory \cite{SS}  
                                            &-7.61509174 &0.2592\\
3-body $\hat H_2$, $M$=0.381$m_e$, exact theory ($N$=16)  
                                            &-1.174475   &1.675828\\
4-body $H_2$, $r_c$=0.01125 a.u., var. theory \cite{SS}
                                            &-1.174474   &1.4011\\
$H_2$, experiment                           &-1.174474   &1.4011\\
\hline
\end{tabular}
}
\caption{The total ground state energy $E$ and the internuclear
distance $R$.}
\l{Table1}
\end{center}
\end{table}
\clearpage

\section{Concluding remarks}\l{Concluding}

In this paper we have shown that the restricted three-body Santilli-Shillady 
isochemical model of the hydrogen molecule admits an exact analytic solution
capable of representing the molecular binding energy in a way accurate
to the sixth digit, $E=-1.174475\au$, and the internuclear distance 
$R = 1.675828 \au$, which is about 19.6\% bigger than the conventional 
experimental value, $R_{exper}[H_2]=1.4011 \au$

We should emphasize that the presented exact analytical solution
includes infinite chain fractions. They still need numerical computation
to reach the characteristic values of $H_2^+$ ion like
systems, such as the ground state energy, with the understanding that
these values can be reached with any needed accuracy. 
For example, at the lengths of the chain fractions
$N=100$ and $N=50$ for the angular and radial eigenvalues, one
achieves accuracy of the ground state energy of about $10^{-12}$.

The general (four-body) Santilli-Shillady isochemical model of $H_2$ 
cannot be, apparently, solved exactly, even in Born-Oppenheimer
approximation, so that Ritz variational approach can be applied here
to get the {approximate} values of the ground state energy
and corresponding internuclear distance.

Ritz variational approach is a good instrument to analyze
few-body problems, and restricted $H_2$ molecule is such a system.
It is wellknown that the
variational solution of the ordinary model of $H_2$ molecule
includes rather complicated analytical calculations of the
molecular integrals, with the hardest part of work
being related to the exchange integral.
Particularly, evaluated exchange integral for $H_2$ molecule
is expressed in terms of a special function (Sugiura's result, 1927).
It is quite natural to expect that even more complications will
arise when dealing with the Hulten potential.

The reason to consider the general four-body Santilli-Shillady model
of $H_2$ molecule, after the made analysis of $H_2^+$ like system
approximate approach to it, is that the stable point-like isoelectronium
$\hat H_2$ model-based theoretical
prediction does not meet the experimental data on $H_2$ molecule,
for the "bare" isoelectronium mass $M=2m_e$, although we achieved 
essentially exact representation of the binding energy taking 
$M=0.308381m_e$. 
Also, this stable point-like isoelectronium (three-body) model 
does not account for essential effect existing in the general 
(four-body) model.
This effect is related to the {potential barrier} between the
region associated to the attractive Hulten potential,
$r_{12}<r_0$, and the region associated to repulsive Coloumb
potential, $r_{12}>r_0$, where $r_0$ is the distance between the
electrons at which Hulten potential is equal to Coloumb potential,
$V(r_{12})=0$; see Eq.(\ref{U}). Characteristics of the barrier
can be extracted from the function $V(r_{12})$.
The barrier is finite for the used values of the parameters 
$V_0$ and $r_c$ so that the electrons penetrate it.
The two $1s$ electrons are thus simultaneously in two regimes, the first is
strongly correlated regime due to short-range attractive Hulten potential
(isoelectronium) and the second is weakly correlated regime
due to the ordinary Coloumb repulsion. Also, there exist a transient
regime corresponding to the region about the equilibrium point,
$r_{12}\simeq r_0$, i.e. inside the barrier.
Schematically, one could thought of that the electrons are, for instance,
10\% in the isoelectronium regime, 1\% in the transient regime, and 89\%
in the Coloumb regime. We stress that in the three-body approach to $H_2$
molecule considered in this paper we have 100\% for the isoelectronium
regime.

Numerical computation by Santilli and Shillady \cite{SS}
based on Gaussian transform techniques and {SASLOBE} computer program
has shown {\it excellent} agreement of the general four-body model 
with experimental data on $H_2$ molecule. They used {\it Gaussian screened 
Coloumb potential} as an approximation to the Hulten potential.
It would be instructive to use Ritz variational approach,
which deals with {\it analytical} calculations, in studying
the four-body Santilli-Shillady model of $H_2$ molecule. One can try it
first for the Gaussian screened Coloumb potential, or exponential screened 
Coloumb potential in which case there is a hope to achieve exact analytical 
evaluation of the Coloumb and exchange integrals.
Being a different approach, this would give a strong support
to the numerical results on the ground state energy obtained
by Santilli and Shillady. Also, having analytical set up
one can make qualitative analysis of the four-body Santilli-Shillady
model of $H_2$ molecule. However, we should to note that these potentials, 
being approximations to the Hulten potential, will yield some  approximate 
models, with corresponding approximate character of the results.

\newpage
\section*{Appendix}

We use $N=16$ power degree approximation, the polynomials $Q_N^{(x)}$
and $Q_N^{(y)}$, to find both
the radial, $\lambda^{(x)}(p,a)$, and angular, $\lambda^{(y)}(p,b)$,
eigenvalues of the {\sc csf}. $Q_N$'s are obtained by the use of
recurrence relations (\ref{Qk}) and definitions of the coefficients
$\alpha_s$, $\beta_s$, $\gamma_s$, $\rho_s$, $\kappa_s$, and $\delta_s$,
where we put $b=0$, i.e. $Z_1=Z_2=1$, and quantum number $m=0$.
Each of the two polynomials has 16 roots for $\lambda$ from which we
select one root which is appropriate due to its asymptotic behavior
at $R\to 0$.
Numerical solution of the equation $\lambda^{(x)}(p,a)=\lambda^{(y)}(p,b)$
gives us the list of values of the electronic 
ground state energy $E(R)=E_{1s\sigma}(R)$, which corresponds
to $1s\sigma_g$ term of the $H_2^+$ ion, as a function of the distance
$R$ between the nuclei. Table~\ref{Table2} presents the result, where
no interpolation has been used. Numerical computation of each point in 
Table~\ref{Table2} took about 88 sec on ordinary Pentium desktop computer.

Below, we present some useful numerical values enabling one to convert 
atomic units, at which $m_e=e=\hbar=1$, to the other units.
Note also that for the energy $1 \au \equiv 1 \hartree$, 
and for the length $1 \au \equiv 1 \bohr$. 
\begin{center}
{\bf Atomic units in terms of the other units}
\end{center}
\begin{center}
\begin{tabular}{|l||l|l|}
\hline
1 a.u. of mass, $m_e$ & 9.10953$\cdot 10^{-28}$& gramms \\
\hline
1 a.u. of charge, $e$ & 1.60219$\cdot 10^{-19}$& Coloumbs \\
\hline
1 a.u. of action, $\hbar$  &  1.05459$\cdot 10^{-27}$ & ${\rm erg\cdot sec}$ \\
\hline
1 a.u. of length, $\frac{\hbar^2}{m_ee^2}$  & 0.529177$\cdot 10^{-8}$& cm \\
\hline
1 a.u. of energy, $\frac{m_ee^4}{\hbar^2}$ & 27.2116& eV \\
\hline
1 a.u. of time, $\frac{\hbar^3}{m_ee^4}$ & 2.41888$\cdot 10^{-17}$& sec \\
\hline
1 a.u. of velocity, $e^2/\hbar$ & 2.18769$\cdot 10^{8}$ & cm/sec\\
\hline
$\alpha =\frac{e^2}{\hbar c}$ & 1/137.0388 & \\
\hline
\end{tabular}
\end{center}

\medskip
\begin{center}
{\bf Conversion of the energy units}
\end{center}
\begin{center}
\begin{tabular}{|c||c|c|c|c|}
\hline
     &  a.u. & eV   & Kcal$\cdot$mole  & cm$^{-1}$      \\
\hline
a.u. &    1     & 27.212  &6.2651$\cdot 10^2$ &2.1947$\cdot 10^5$ \\
\hline
eV   & 3.6749$\cdot 10^{-2}$& 1 & 23.061   &  8065.48 \\
\hline
Kcal$\cdot$mole &1.5936$\cdot 10^{-3}$ & 4.3364$\dot 10^{-2}$& 1&3.4999$\cdot 10^2$ \\
\hline
cm$^{-1}$& 4.5563 $\cdot 10^{-6}$ &1.2398$\dot 10^{-4}$& 2.8573$\cdot 10^{-3}$&1\\
\hline
\end{tabular}
\end{center}

\clearpage

\begin{table}[ht]
\begin{center}
{\scriptsize
\begin{tabular}{ll|ll}
\cline{1-4}\\
 {\small $R$, a.u.} &{\small $E(R)$, a.u.}
&{\small $R$, a.u.} &{\small $E(R)$, a.u.}\\
\cline{1-4}\\
0.0& -1.99999999761099225          &2.01& -1.10013870349441877\\
0.1& -1.97824134920757046          &2.05& -1.09030214496519061\\
0.2& -1.92862028526774320          &2.1& -1.07832542203506842\\
0.3& -1.86670393395684293          &2.2& -1.05538508113994433\\
0.4& -1.80075405253452878          &2.3& -1.03371349485820318\\
0.5& -1.73498799160719041          &2.4& -1.01322030525887973\\
0.6& -1.67148471440012302          &2.5& -0.99382351101203490\\
0.7& -1.61119720301672586          &2.6& -0.97544858094023219\\
0.8& -1.55448006449772595          &2.7& -0.95802766004904907\\
0.9& -1.50138158334624467          &2.8& -0.94149886061738322\\
1.0& -1.45178630031448951          &2.9& -0.92580563147803989\\
1.1& -1.40550252191841256          &3.0& -0.91089619738235434\\
1.2& -1.36230785783351171          &3.1& -0.89672306127076382\\
1.3& -1.32197139010318509          &3.2& -0.88324255989570446\\
1.4& -1.28426925496894185          &3.3& -0.87041447461742614\\
1.5& -1.24898987186705512          &3.4& -0.85820167794222435\\
1.6& -1.21593722446146546          &3.5& -0.84656982450508629\\
1.7& -1.18493139974611416          &3.6& -0.83548707392472181\\
1.8& -1.15580915764590441          &3.7& -0.82492384412924800\\
1.9& -1.12842156954614458          &3.8& -0.81485259165546253\\
1.95& -1.11533575206408963         &3.9& -0.80524772550750985\\
1.99& -1.10514450160298682         &4.0& -0.79608496995054425\\
2.0& -1.10263415348745197          &8.0& -0.62757022044109352\\
\cline{1-4}\\
\cline{1-4}
\end{tabular}
}
\caption{The electronic energy of $H_2^+$ ion (see Fig.~1).}
\l{Table2}
\end{center}
\end{table}

Minimum of the energy $E_{1s\sigma}(R)$ is
$E_{1s\sigma}= -1.9999999976 \au$ at $R=0$, which reproduces the known value
$E_{1s\sigma}=-2 \au$ to a very high accuracy. Moreover, one can compare
Table~\ref{Table2} and the table of Ref.\cite{Wind} to see that 
each energy value in Table~\ref{Table2} does reproduce Wind's result 
up to seven decimal places. This means that our numerical calculations 
are correct.

We remark that Wind used $N=50$ approximation and presented seven 
decimal places while we use $N=16$ approximation and present seventeen 
decimal places. Alas, there is no need to keep such a high accuracy, 
and also Wind mentioned that even $N=10$ approximation gives the same 
result, up to seven digits. 

By making 16th-order interpolation of the points in Table~\ref{Table2}
and adding to it the potential of interaction between the nuclei, $1/R$, 
we obtain the list of values of the total energy presented in 
Table~\ref{Table3}.
It reveals the only minimum of the total energy,
$E(R)+R^{-1}=E_{min}=-0.6026346 \au$ at the distance 
$R=R_{opt}=1.9971579 \au$

\begin{table}[ht]
\begin{center}
{\scriptsize
\begin{tabular}{ll|ll}
\cline{1-4}\\
 {\small $R$, a.u.} &{\small $E(R)+R^{-1}$, a.u.}
&{\small $R$, a.u.} &{\small $E(R)+R^{-1}$, a.u.}\\
\cline{1-4}\\
0.1 &+8.02176    &2.1 &-0.602135\\
0.2 &+3.07138    &2.2 &-0.600840\\
0.3 &+1.46663    &2.3 &-0.598931\\
0.4 &+0.69924    &2.4 &-0.596554\\
0.5 &+0.26501    &2.5 &-0.593824\\
0.6 &-0.004818   &2.6 &-0.590833\\
0.7 &-0.182626   &2.7 &-0.587657\\
0.8 &-0.304480   &2.8 &-0.584356\\
0.9 &-0.390270   &2.9 &-0.580978\\
1.0 &-0.451786   &3.0 &-0.577563\\
1.1 &-0.496412   &3.1 &-0.574142\\
1.2 &-0.528975   &3.2 &-0.570743\\
1.3 &-0.552741   &3.3 &-0.567384\\
1.4 &-0.569984   &3.4 &-0.564084\\
1.5 &-0.582323   &3.5 &-0.560856\\
1.6 &-0.590937   &3.6 &-0.557709\\
1.7 &-0.596696   &3.7 &-0.554654\\
1.8 &-0.600254   &3.8 &-0.551695\\
1.9 &-0.602106   &3.9 &-0.548837\\
2.0 &-0.602634   &4.0 &-0.546085\\
\cline{1-4}\\
\cline{1-4}
\end{tabular}
}
\caption{The total energy of $H_2^+$ ion (see Fig.~2).}
\l{Table3}
\end{center}
\end{table}

The results collected in Table~\ref{Table4} have been obtained directly
by numerical calculations with the use of replacements
$p^2 \to 2p^2$ and $a\to 4a$,
where $p$ and $a$ are defined by Eq.(\ref{pab}), in the coefficients
$\alpha_s$, $\beta_s$, $\gamma_s$, $\rho_s$, $\kappa_s$, and $\delta_s$
of the recurrence relations.
These replacements have been made due to Eq.(\ref{ca}),
with the mass parameter $M=2$ and the charge parameter $q=-2$,
corresponding to the stable point-like isoelectronium of mass $M=2m_e$ and 
charge $-2e$.
In addition, it turns out that Table~\ref{Table4} can be derived 
directly from Table~\ref{Table2} by the use of rescalements
$R\to R/4$ and $E\to 8E$.
This remarkable property is confirmed by the scaling method developed in
Sec.~\ref{TheScaling}, and proves that the scaling method is correct.
By adding $1/R$ to the isoelectronic energy values of Table~\ref{Table4}
we obtain Table~\ref{Table5} 
showing the total energy of the $\hat H_2$ system, at the mass
$M=2m_e$. The minimum of the total energy is found 
$E(R)+R^{-1}=E_{min}=-7.617041 \au$ at 
$R=R_{opt}=0.258399 \au$


\clearpage
\newpage
\begin{table}
\begin{center}
{\scriptsize
\begin{tabular}{ll|ll}
\cline{1-4}\\
 {\small $R$, a.u.} & {\small $E(R)$, a.u.}
&{\small $R$, a.u.} & {\small $E(R)$, a.u.}\\
\cline{1-4}\\
0.00&-16.0000000000000008  &0.50&-8.82107371596060652\\
0.05&-15.4289613540288446  &0.55&-8.44308064942978475\\
0.10&-14.4060324481253427  &0.60&-8.10576243870545276\\
0.15&-13.3718774940028106  &0.65&-7.80358864752256309\\
0.20&-12.4358407555134897  &0.70&-7.53199088488360768\\
0.225&-12.0110526341241952 &0.75&-7.28716957910914597\\
0.25&-11.6142894041116995  &0.80&-7.06594047783391143\\
0.275&-11.2440221722385613 &0.85&-6.86561342351093717\\
0.30&-10.8984629511501962  &0.90&-6.68389659171409977\\
0.35&-10.2741538838779300  &0.95&-6.51882073404630535\\
0.40&-9.72749779604447084  &1.00&-6.36867910416260141\\
0.45&-9.24647351730713928  &    &\\
\cline{1-4}\\
\cline{1-4}
\end{tabular}
}
\caption{The isoelectronium energy of the three-body $\hat H_2$ system,  
at the mass $M=2m_e$ (see Fig.~3).}
\l{Table4}
\end{center}
\end{table}
\nopagebreak[4]

\begin{table}
\begin{center}
{\scriptsize
\begin{tabular}{ll|ll}
\cline{1-4}\\
 {\small $R$, a.u.} & {\small $E(R)$, a.u.}
&{\small $R$, a.u.} & {\small $E(R)$, a.u.}\\
\cline{1-4}\\
0.010&+84.0273769895390465 &0.150&-6.70521082733614370\\
0.015&+50.7305619003569852 &0.200&-7.43584075551348888\\
0.020&+34.1124961901906909 &0.250&-7.61428940411169996\\
0.025&+24.1713076006601701 &0.300&-7.56512961781686321\\
0.030&+17.5720885190207170 &0.350&-7.41701102673507239\\
0.035&+12.8849357295675638 &0.400&-7.22749779604447084\\
0.040&+9.39442987714794597 &0.450&-7.02425129508491696\\
0.045&+6.70277469374505585 &0.500&-6.82107371596060652\\
0.050&+4.57103864597115538 &0.550&-6.62489883124796641\\
0.055&+2.84698239022676524 &0.600&-6.43909577203878669\\
0.060&+1.42896579635860376 &0.650&-6.26512710906102388\\
0.065&+0.24650856005471055 &0.700&-6.10341945631217797\\
0.070&-0.75082137484248256 &0.750&-5.95383624577581205\\
0.075&-1.60007720159364552 &0.800&-5.81594047783391143\\
0.080&-2.32910343012903808 &0.850&-5.68914283527564279\\
0.085&-2.95923562411244667 &0.900&-5.57278548060298906\\
0.090&-3.50710131449860362 &0.950&-5.46618915509893721\\
0.095&-3.98585308299203866 &1.000&-5.36867910416260141\\
0.100&-4.40603244812534278 &     &\\
\cline{1-4}\\
\cline{1-4}
\end{tabular}
}
\caption{The total energy of the three-body $\hat H_2$ system,  
at the mass $M=2m_e$ (see Fig.~4).}
\l{Table5}
\end{center}
\end{table}

\clearpage
\pagebreak[4]
\begin{table}
\begin{center}
{\scriptsize
\begin{tabular}{l|l|l}
\cline{1-3}\\
{\small $M$, a.u.} & {\small $E_{min}(M)$, a.u.}
&{\small $R_{opt}(M)$, a.u.}\\
\cline{1-3}\\
0.10 &-0.380852 &5.167928\\
0.15& -0.571278 &3.445291\\
0.20 &-0.761704 &2.583964\\
0.25&-0.952130 &2.067171\\
0.26&-0.990215&1.987664\\
0.27&-1.028300&1.914050\\
0.28&-1.066385&1.845688\\
0.29&-1.104470&1.782044\\
0.30&-1.142556&1.722645\\
0.307&-1.169215&1.683367\\
0.308&-1.173024&1.677899\\
0.308381&-1.174475&1.675828\\
0.309&-1.176832&1.672471\\
0.31&-1.180641&1.667073\\
0.32&-1.218726&1.614977\\
0.33&-1.256811&1.566041\\
0.34&-1.294896&1.519981\\
0.35&-1.332982&1.476553\\
0.40&-1.523408&1.291982\\
0.45&-1.713834&1.148428\\
0.50&-1.904260&1.033585\\
0.75&-2.856390&0.689058\\
1.00&-3.808520&0.516792\\
1.25&-4.760650&0.413434\\
1.50&-5.712780&0.344529\\
1.75&-6.664910&0.295310\\
2.00&-7.617040&0.258396\\
\cline{1-3}\\
\cline{1-3}
\end{tabular}
}
\caption{The minimal total energy $E_{min}$ and the optimal internuclear 
distance $R_{opt}$ of the three-body $\hat H_2$ system as functions of the 
mass $M$ of the stable point-like isoelectronium (see Figs.~5--8).}
\l{Table6}
\end{center}
\end{table}
Table~\ref{Table6} presents result of calculations of the {\it minimal}
total energies and corresponding {\it optimal} distances, at various 
values of the isoelectronium mass parameter $M=\eta m_e$ 
($M=\eta$, in atomic units).
We have derived some 27 tables (such as Table~\ref{Table5}) 
from Table~\ref{Table2} by the scaling method according to 
Eq.(\ref{scal2}), and find minimum of the total energy in 
each table, together with the corresponding optimal distance. 
Then we collected
all the obtained energy minima and optimal distances in Table~\ref{Table6}. 
With the fourth order interpolation/extrapolation, the graphical 
representations of Table~\ref{Table6} show 
(see Figures~5--8)
that the minimal total energy behaves as $E_{min}(M) \simeq -3.808M$,
and the optimal distance behaves as $R_{opt}(M)\simeq 0.517/M$, 
to a good accuracy. One can see that at $M=2m_e$ we have 
$E_{min}(M)=-7.617040 \au$ and $R_{opt}(M)=0.258396 \au$, which recover 
the earlier obtained values $E_{min}=-7.617041 \au$ and 
$R_{opt}=0.258399 \au$ 
of Table~{\ref{Table5}}, to a high accuracy, thus showing once again 
correctness of the used scaling method. 
In fact, the values of $E$ and $R$ for $M=1.50m_e$, $M=1.75m_e$, 
and $M=2.00m_e$ in Table~{\ref{Table6}} have been obtained by extrapolation 
so they are not as much accurate as they are in Table~{\ref{Table5}}. 
However, this is not of much importance here because we use them only to 
check the results of the scaling method.

The main conclusion following from Table~{\ref{Table6}} is that
the mass parameter value $M=0.308381m_e$ fits the energy value 
$E_{min}(M)=-1.174475 \au$, with the corresponding 
$R_{opt}(M)=1.675828 \au$, which appears to be about
19.6\% bigger than the experimental value $R_{exper}[H_2]=1.4011 \au$
The total energy as a function of internuclear distance, for this value 
of mass, $M=0.308381m_e$, is shown in Figure~9 
to illustrate that the obtained optimal distance $R_{opt}=1.675828 \au$ 
corresponds to a minimal value of the total energy. 

\clearpage
\pagebreak[4]

\begin{figure}[t]
\plot{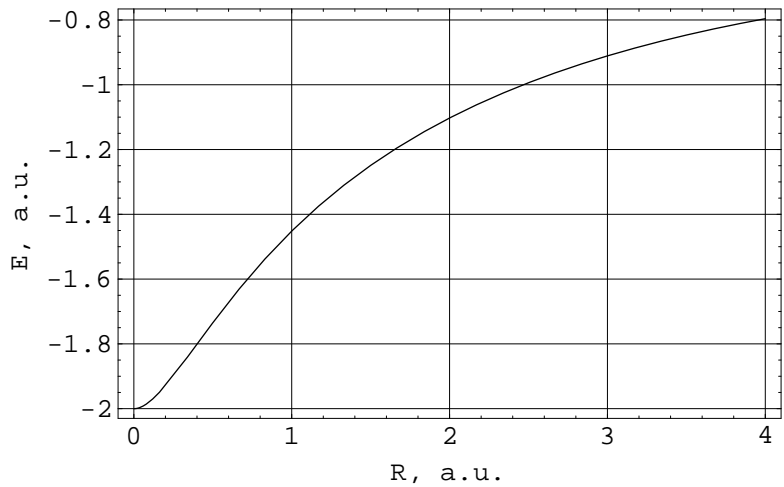}
\caption{The electronic energy $E(R)$ of $H_2^+$ ion as a function of 
the internuclear distance $R$.}
\label{Fig1}
\end{figure}

\begin{figure}[b]
\plot{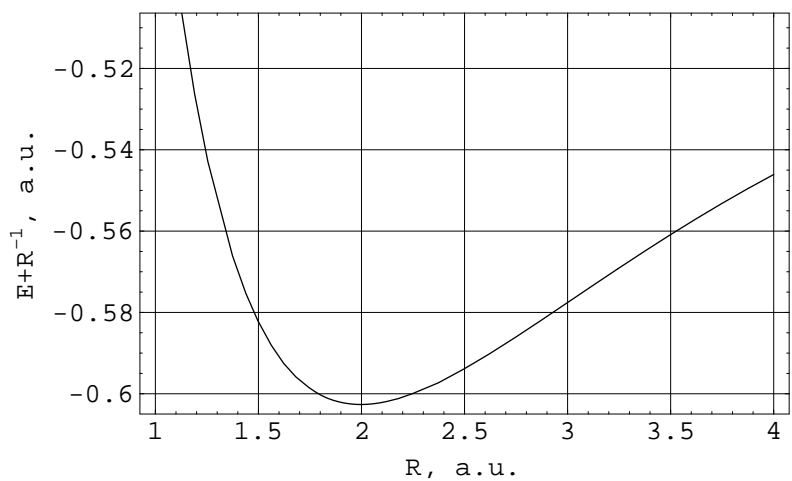}
\caption{The total energy $E(R)+R^{-1}$ of $H_2^+$ ion as a function of 
the internuclear distance $R$.}
\label{Fig2}
\end{figure}

\clearpage

\begin{figure}
\plot{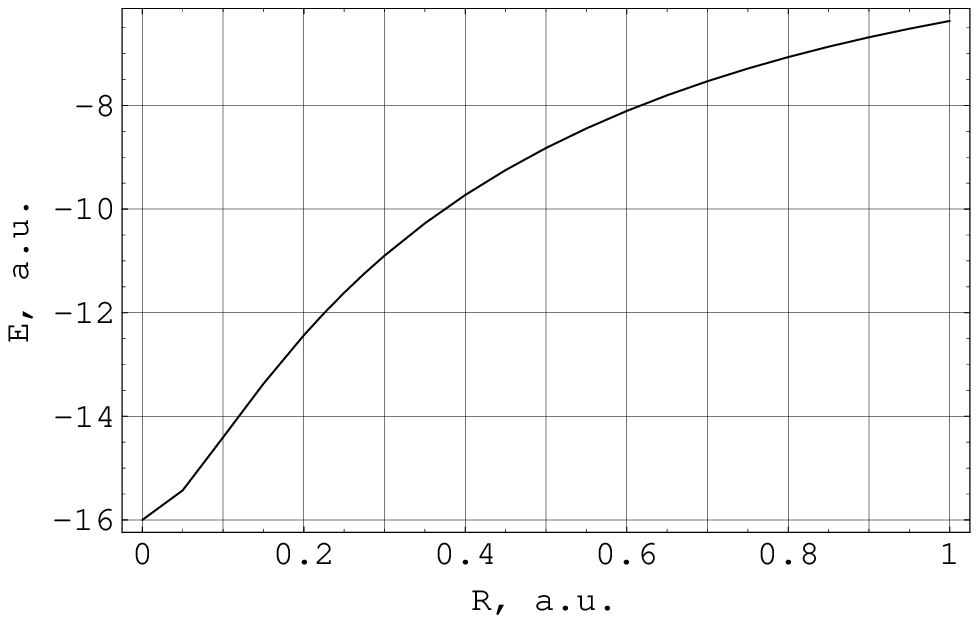}
\label{Fig3}
\caption{The isoelectronium energy $E(R)$ of the $\hat H_2$ system as 
a function of the internuclear distance $R$, at the isoelectronium 
mass $M=2m_e$.}
\end{figure}

\begin{figure}
\plot{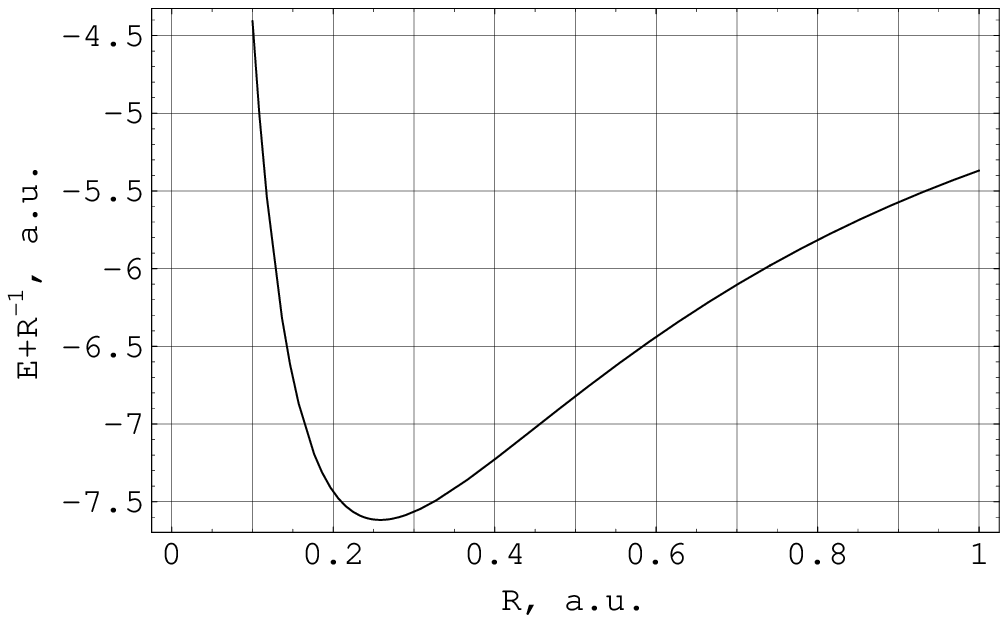}
\label{Fig4}
\caption{The total energy $E(R)+R^{-1}$ of the $\hat H_2$ system 
as a function of the internuclear distance $R$, at the isoelectronium 
mass $M=2m_e$.}
\end{figure}

\clearpage

\begin{figure}
\plot{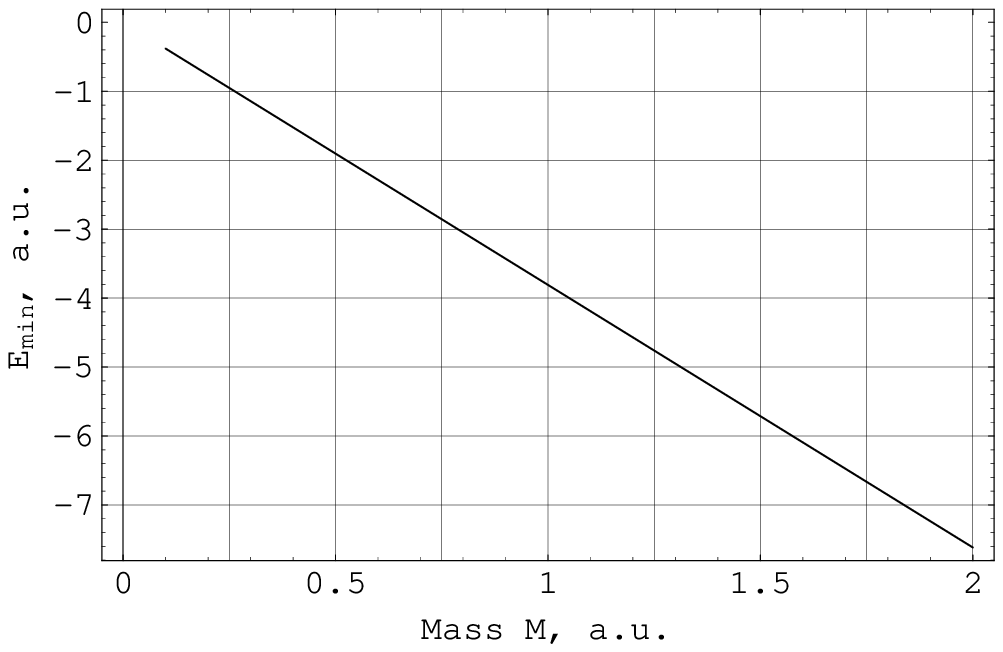}
\label{Fig5}
\caption{The minimal total energy $E_{min}(M)$ of the $\hat H_2$ system as 
a function of the isoelectronium mass $M$.}
\end{figure}

\begin{figure}
\plot{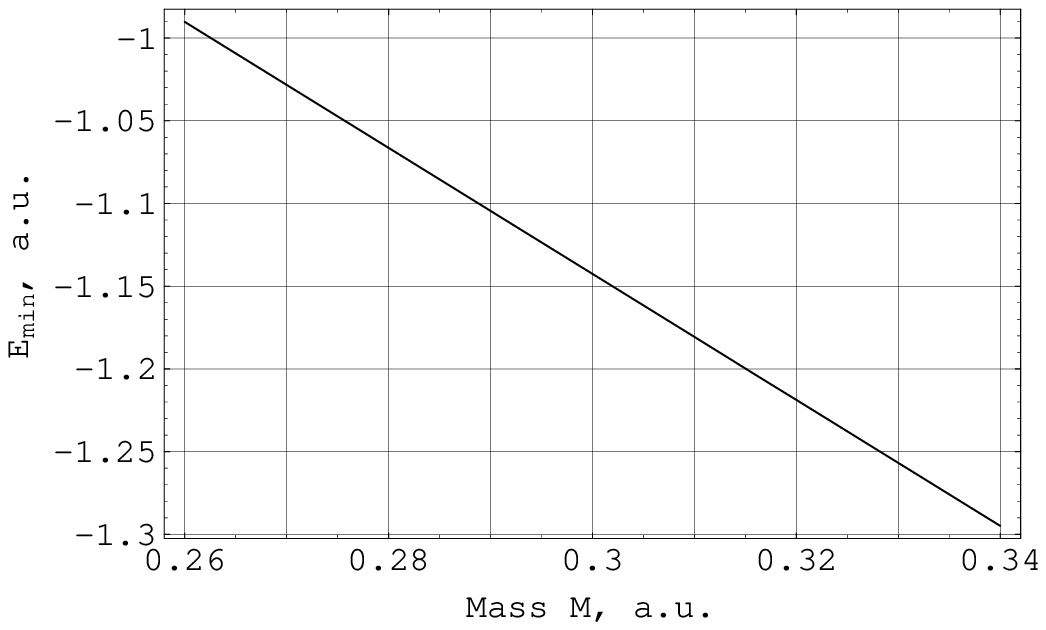}
\label{Fig6}
\caption{The minimal total energy $E_{min}(M)$ of the $\hat H_2$ system as 
a function of the isoelectronium mass $M$. More detailed view.}
\end{figure}

\clearpage

\begin{figure}
\plot{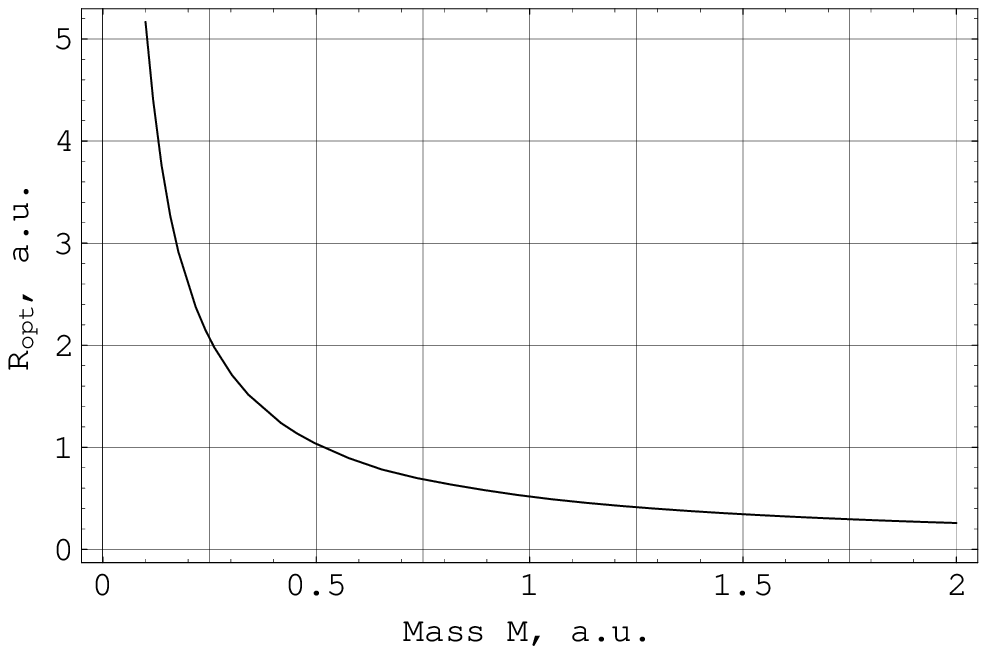}
\label{Fig7}
\caption{The optimal internuclear distance $R_{opt}(M)$ of the $\hat H_2$ 
system as a function of the isoelectronium mass $M$.}
\end{figure}

\begin{figure}
\plot{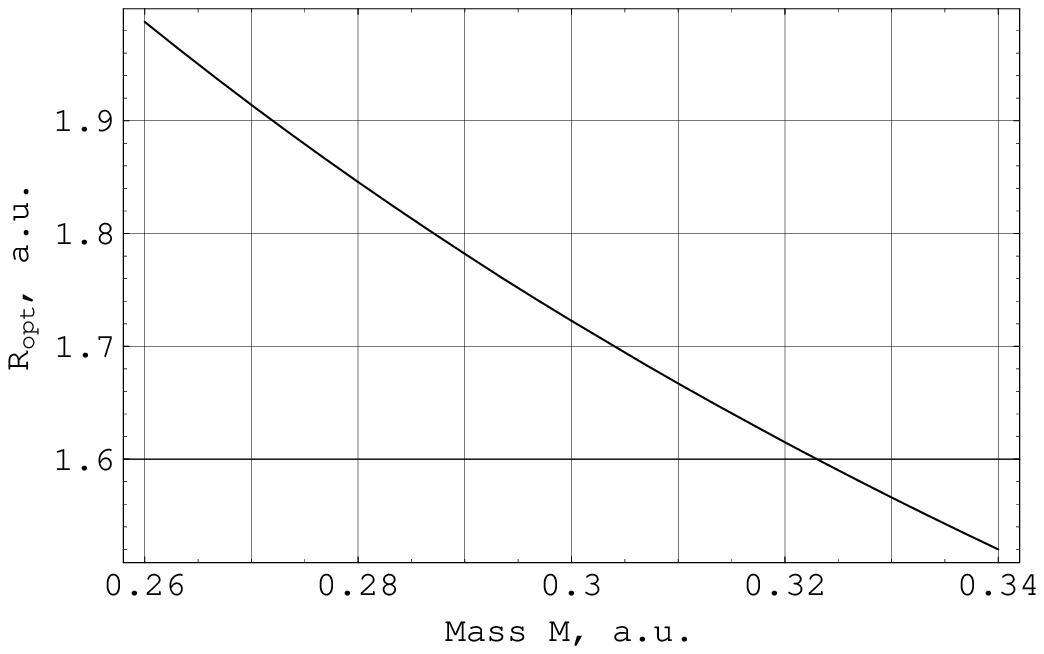}
\label{Fig8}
\caption{The optimal internuclear distance $R_{opt}(M)$ of the $\hat H_2$ 
system as a function of the isoelectronium mass $M$. More detailed view.}
\end{figure}

\clearpage

\begin{figure}
\plot{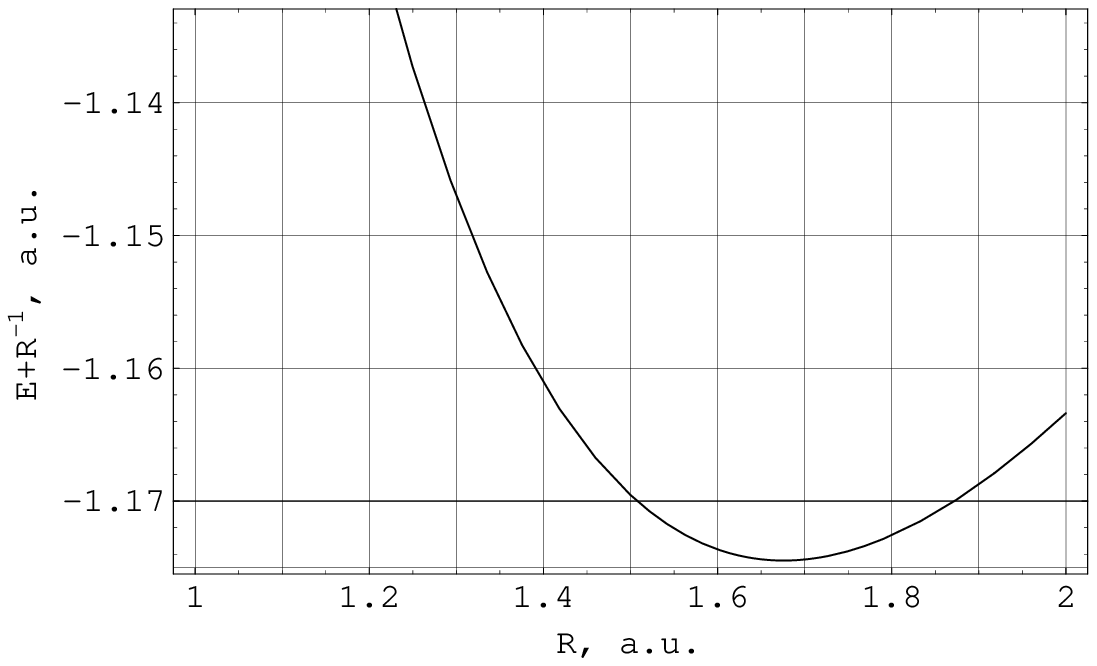}
\label{Fig9}
\caption{The total energy $E(R)+R^{-1}$ of the $\hat H_2$ system 
as a function of the internuclear distance $R$, at the isoelectronium 
mass $M=0.308381m_e$. }
\end{figure}

\clearpage
\pagebreak[4]

\end{document}